\documentclass[notitlepage,aps,prd,groupedaddress,nofootinbib]{revtex4-1}

\usepackage[sc]{mathpazo} 
\usepackage[T1]{fontenc} 
\usepackage{microtype} 

\usepackage[hmarginratio=1:1,top=32mm,columnsep=20pt]{geometry} 
\usepackage[hang, small,labelfont=bf,up,textfont=it,up]{caption} 
\usepackage{booktabs} 
\usepackage{float} 
\usepackage{hyperref} 
\usepackage{graphics}
\usepackage{graphicx}
\usepackage{float}
\usepackage{amssymb}
\usepackage{slashed}
\usepackage{fancybox}
\usepackage{amsfonts}
\usepackage{epsfig}
\usepackage{rotating}
\usepackage{amsmath}
\usepackage{amsmath}%
\usepackage{wasysym}%
\usepackage{braket}
\usepackage{epsfig}
\usepackage{lettrine} 
\usepackage{paralist} 

\usepackage{fancyhdr} 
\allowdisplaybreaks

\begin{document}


\title{Unified dark matter with intermediate symmetry breaking scales}


\author{Stephen J. Lonsdale}
\email[]{Corresponding author: lsj@student.unimelb.edu.au}
\affiliation{ARC Centre of Excellence for Particle Physics at the Terascale, School of Physics,\\ The University of Melbourne, Victoria 3010, Australia}


\begin{abstract}
Asymmetric symmetry breaking models dynamically break the $G \times G$ gauge symmetries of mirror models to distinct subgroups in the two sectors. The coincidental abundances of visible and dark matter, $\Omega_{DM} \simeq 5\Omega_{VM}$, motivates
asymmetric dark matter theories where similar number densities of baryons in each sector are explained by their connected origins. However the question of why the baryons of two sectors should have similar mass remains. 
In this work we develop an alternative class of asymmetric symmetry breaking models which unify the dark and visible 
sectors while generating a small difference in the mass scale of the baryons of each sector. By examining the different paths that the SO(10) GUT group can take in breaking to gauge symmetries containing SU(3) 
we can adapt the mechanism of asymmetric symmetry breaking to demonstrate models in which originally unified visible and dark sectors have isomorphic color gauge groups at low energy yet pass 
through different intermediate gauge groups at high energy. Through this, slight differences in the running coupling evolutions and thus the confinement scales of the two sectors are generated.
\end{abstract}
\maketitle 

\thispagestyle{fancy} 



                  \section{ \bf Introduction}\label{sec:Intro}
                  The present understanding of our universe places all of the matter that we understand as just a small fraction of the total amount of matter and energy that make up the cosmos.
                  This visible matter (VM), made up of the particles of the standard model interacting under the gauge forces described by the group $SU(3)\times SU(2)\times U(1)$ accounts for only 4.9 percent of the total mass-energy 
                  while the remainder is made up of the currently unknown dark matter (DM), with 26.8 percent, and dark energy which drives the acceleration of the universe accounting for 68.3 percent.
                  The similarity in the abundances of visible and dark matter suggests a common origin, since if the two forms of matter were completely independent their relative abundances would likely be dissimilar.
                  Asymmetric dark matter (ADM) models seek to explain this observed ratio of visible and dark matter of $\Omega_{DM} \simeq 5\Omega_{VM}$ by 
                  the conservation of a single global quantum number. By establishing a symmetry in a linear combination of baryon and dark baryon numbers, 
                  the matter-antimatter annihilations and chemical reprocessing taking place result in a relation between the number densities of visible and dark baryons \cite{Davoudiasl:2012uw, Petraki:2013wwa, Zurek:2013wia}.
                  Though armed with an explanation for similar baryon number densities, ADM models are still in need of an explanation as to why visible and dark baryons should have similar mass.
                  In \cite{Lonsdale:2014wwa} the mechanism of  \emph{asymmetric symmetry breaking (ASB)} was introduced to develop a way of connecting these masses by generating an SU(3) confinement 
                  scale in each sector which is similar but slightly different due to a grand unification of the two sectors for which the originally unified $G_V \times G_D$ gauge group is broken differently for each sector. This was used in the context of an 
                  $SU(5) \times SU(5)$ model where the hidden sector was broken to a dark $SU(3)$.
                  Such a model then allows for variations in the masses of fermions in each sector and alters the running couplings of the surviving SU(3) gauge symmetries that confine the visible and dark baryons. 
                  This model worked by  using the fact that since the two sectors unify into a single $G \times G$ group at high energy with a $\Bbb{Z}_2$ mirror symmetry, the values of the coupling constants at the GUT scale are the same and  different 
                  quark masses of the two sectors differentiate the evolution of the couplings slightly to produce confinement scales of similar but distinct value.
                
                  In this work we explore the ability of spontaneous symmetry breaking to generate similar results from different GUT breaking chains in the two sectors in an $SO(10) \times SO(10)$ 
                  theory. These different gauge symmetry breaking chains can result from a simple extension of the mechanism of ASB and allows one to create regions where the coupling evolution differs 
                  in the two sectors without considering fermion mass generation.
                  The models that we explore here are larger extensions to mirror symmetric models which have been explored in many contexts \cite{Lee:1956qn,Kobzarev:1966,Pavsic:1974rq,Blinnikov:1982eh,Blinnikov:1983,Foot:1991bp,Foot:1991py,Foot:1995pa,Berezhiani:1995yi,Foot:2000tp,Berezhiani:2000gw,Ignatiev:2003js,Foot:2003jt,Foot:2004pq,Berezhiani:2003wj,Ciarcelluti:2004ik,Ciarcelluti:2004ip,Foot:2014mia}, 
                  where in this work the mirror symmetry serves only at high energy and the low energy features of the two sectors can be vastly different. We use this to develop a way of explaining the similarity of 
                  DM mass, the focus of this paper. In further work it would be interesting to see more complete theories that explore the baryogenesis of the two sectors such as in the recent work of \cite{Gu:2014nga} where an $SO(10) \times SO(10)$ model 
                  explored baryogenesis via leptogenesis with visible and dark QCD scales set at similar values.
                  Our work is also related to other investigations into the possibility of a dark QCD such as \cite{Bai:2013xga,Barr:2013tea,Ma:2013nga,Newstead:2014jva,Boddy:2014yra,Yamanaka:2014pva, Higaki:2013vuv}.
                  
                  The next section will review the motivation for such models by examining how the running of coupling constants in gauge theories with unification can be 
                  used to link the color confinement scales of the two sectors.
                  From there Section \ref{sec:SO(10)} will discuss SO(10) models and their appeal as the choice of GUT group to be implemented with this method.
                  Following this Section \ref{sec:Breakchains} and Section \ref{sec:ASB} will discuss the paths of symmetry breaking that we can take within SO(10) models and 
                  how these can be used in asymmetric symmetry breaking models to create the SM in one sector with an SU(3) group in the dark sector. 
                  We will then move on to Section \ref{sec:SUSY} where we will explore similar models within the supersymmetric framework, while Section \ref{sec:Results} will examine the results from a broad 
                  range of these possible scenarios and their effect on the dark QCD scale. 
                  Finally in Section \ref{sec:Pheno} we will discuss the constraints on some of these models and the outlook for such theories.
 
                  \section{Dimensional transmutation}\label{sec:DT}
                  Our objective is to develop $SO(10)\times SO(10)$ models that can account for the similarity in mass of visible and dark matter. The overwhelming majority of the mass of visible matter comes from dimensional transmutation where a dimensionful 
                  parameter is created at the scale at which a coupling begins to diverge and the theory becomes non-perturbative. The masses of the protons and neutrons which dominate the visible sector in the present universe come from the confinement scale of QCD where the
                  coupling constant of the color force becomes large at low energy. This feature of asymptotically free theories presents an elegant way to introduce mass scales into a theory. The capacity to yield such scales at low energy comes from the
                  negative sign of the beta function of a non-Abelian gauge theory. The running coupling evolution is described by the logarithmic dependence on energy scale,
                  \begin{equation}
                  \alpha_s(\mu) = \frac{\alpha_s(\mu_0)}{1 - (b_0/4\pi)\alpha_s(\mu_0)\ln(\mu^2/{\mu_0}^2)},
                  \end{equation}
                  such that at low energy scales the value of $\alpha_s$ grows exponentially. This asymptote sets the energy scale of the proton mass after chiral symmetry breaking when colored particles are confined to bound states. 
                  In a general non-Abelian gauge theory for group G the beta function at one loop is given by    
                  \begin{equation}
                 \beta(g)_{(1\;Loop)} = \frac{g^3}{16 \pi^2}  (-\frac{11}{3}R_{Gauge} + \frac{4}{3}R_{Dirac} + \frac{2}{3}R_{Majorana} + \frac{2}{3}R_{Weyl} + \frac{1}{3}R_{C.Scalar} + \frac{1}{6}R_{Scalar}),
                  \end{equation}
                  where $\beta(g)_{(1\;Loop)}=\frac{g^3}{16 \pi^2}b_0$ and the factors of R are the indices for the choice of multiplet(m) defined as
                  \begin{equation}
                   Tr(t^a t^b)=\delta^{ab} \times R(m),
                  \end{equation}
                  and are calculated for each copy  of the gauge fields, which are necessarily in the adjoint representation of G, followed by the Dirac, Majorana, and Weyl fermions and finally complex and real scalars. 
                  For the familiar QCD group SU(3), 
                  the beta function becomes
                  \begin{equation}
                  b_0 = -11 + \frac{2}{3}n_f,
                  \end{equation}
                  with $n_f$ the number of flavors. 
                  In the standard model the coupling of QCD goes non-perturbative at $\approx 200$ MeV. In this paper we seek 
                  to explain the similarity of visible and dark matter masses by assuming that DM similarly gains its mass by dimensional transmutation and that the confinement scales of the two 
                  sectors are linked to each other by their different evolution from a common starting point at the GUT scale.
                  These differences can occur spontaneously from a completely mirror symmetric model thanks to asymmetric symmetry breaking where the absolute minima of the potential are such that 
                  the vacuum structure of each sector is necessarily different. 
                  The goal of this work is to construct a broad outline of the possible models in which a GUT theory with a discrete $\Bbb{Z}_2$ symmetry can naturally explain the similarity of visible and dark matter masses by
                  spontaneously breaking the symmetries of the two sectors through different subgroups while ending with at least one copy of SU(3) in each sector. In this manner the 
                  confining scale of the dark QCD is related to that of the standard model through the unified couplings at high scale,
                  but within intermediate symmetry breaking scales the coupling constants run differently due to the contribution from the gauge bosons of their respective groups. 
                  It thus becomes effectively the first term in Eq. 2 that changes at particular mass scales allowing for the generation of different confinement scales rather than
                  the second term in Eq. 4 at the quark mass thresholds as in \cite{Lonsdale:2014wwa}. 
                  In this work we will not examine any differences resulting from quark mass thresholds though of course the two effects could be utilized in a single theory. We will focus on those cases where after the altered running of the two QCDs is 
                  established the dark QCD coupling will confine at a higher energy scale as this is more suited to ADM where mass scales of around one order of magnitude higher are compatible. 
                  Figure 1 shows the divergence of the two SU(3) theories after running at different rates for a segment of the high energy regime.
                 
\begin{figure}[t!]
\centering
\includegraphics[angle=0,width=0.7\textwidth]{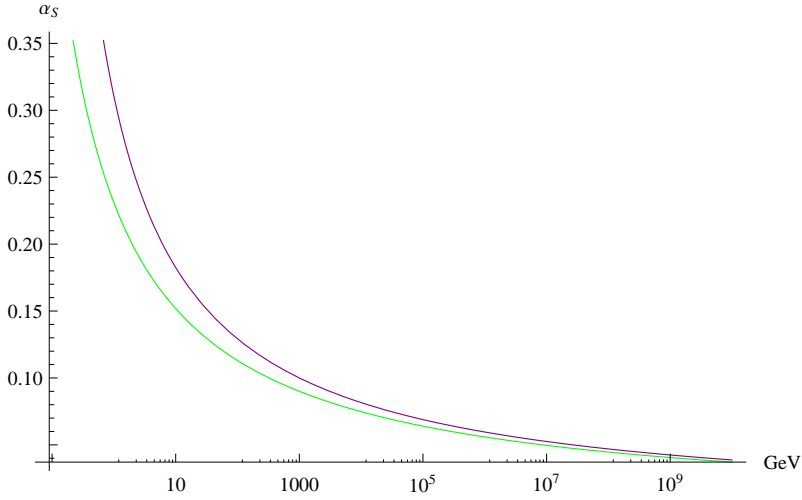}
\caption{The confinement of the dark sector QCD occurs at a higher scale than its visible counterpart after asymmetric symmetry breaking. The top line shows $\alpha_D$ after running as SU(4) for two orders of magnitude at a high energy scale while $\alpha_V$ remains SU(3). }
\end{figure}
                  A number of other works have explored similar concepts of generating the confinement scale of a dark QCD in order to explain the DM mass coincidence. 
                  In particular this work is related to that of \cite{Tavartkiladze:2014lla} where $\Bbb{Z}_2$ symmetric SU(5) GUTs were explored for generating confined states at low scales. 
                  The present work however seeks to expand the technique of asymmetric symmetry breaking beyond SU(5) theories to the SO(10) gauge group and so we move on to a discussion of its features.

                            \section{ \bf SO(10) $\times$ SO(10) Models}\label{sec:SO(10)}
                            The group SO(10) presents an appealing avenue for GUT extensions to the standard model beyond the minimal cases. It has the benefit of allowing each generation of 
                            fermions to fit within a single SO(10) multiplet including the right hand neutrino. Most SO(10) models require at least two Higgs multiplets to break the full symmetry down to the standard model. 
                            Typical choices include one set of fields in \textbf{45} or \textbf{54} representations and another in \textbf{10}, \textbf{16} or \textbf{126} dimensional representations \cite{Ross:1985ai}. 
                            The choice of 126 for the second is appealing as it allows the generation of fermion masses by Yukawa coupling to the 3 copies of $16_f$ which contain the fermions of the standard model.
                            Since there are two multiplets required to break SO(10) to the standard model gauge group, the work of \cite{Lonsdale:2014wwa} can be naturally extended to SO(10) where the visible and dark sectors required two 
                            Higgs representations in each sector to carry out asymmetric symmetry breaking. 
                            By giving a non-zero VEV to all four representations in such a manner that representations paired under the $\Bbb{Z}_2$ symmetry gain VEVs of different sizes, the gauge group of each sector will be different for small segments of the range between the GUT scale 
                            and the low energy theory.
                            The parameter space of this particular type of  model can be quite small as we shall see in Section \ref{sec:ASB} and therefore leads us to consider non-minimal multi step 
                            breaking chains in $SO(10)_V \times SO(10)_D$ models for more than four Higgs multiplets. We are chiefly concerned with paths that can break SO(10) to a gauge sector containing SU(3) in the dark sector while breaking to the SM gauge group in the VS. 
                            Since our primary goal is to generate dark confinement scales only slightly above that of the visible sector, we will limit ourselves to models where this is 
                            the result, that is $\Lambda_D > \Lambda_V$. The case of $\Lambda_D = \Lambda_V$ can also appear, often in the limiting cases where the intermediate scales approach the GUT scale. 
                            
                            To illustrate the concept consider the case where
                            \begin{equation}
                            SO(10)_V  \rightarrow_{M_X}  SU(4) \times SU(2) \times SU(2) \rightarrow_{M_I} SU(3) \times SU(2) \times U(1) ,
                            \end{equation}
                             while in the dark sector 
                            \begin{equation}
                            SO(10)_D  \rightarrow_{M_X}  SU(5) \rightarrow_{M_I} SU(3) \times SU(2) \times U(1).
                            \end{equation}
                            In the visible sector this could be done with with a Higgs multiplet which transforms as a \textbf{54}  and which gains a VEV at the scale $M_X$ while in the dark sector we have a \textbf{45}. 
                            Then a pair of \textbf{16} + $\mathbf{\overline{16}}$ or \textbf{126} + $\mathbf{\overline{126}}$ representations could gain VEVs in both sectors at the scale $M_I$ where each sector becomes standard model-like. The use of a pair of conjugate representations allows for such fields to be included in the superpotential 
                            in supersymmetric theories and also allows us to invoke Michel's conjecture which states that for conjugate pairs such as these, or for real irreducible representations, the symmetry breaking must be to a maximal little group \cite{Michel:1980pc,Slansky:1981yr}.
                            This pair of breaking chains is a particularly simple example where we have only two scales, $M_X$ and $M_I$, however in general it is possible for the intermediate scales of the two sectors to be independent.
                            In such a scenario we have only the distance between the two scales $M_X$ and $M_I$ that determines the size of the difference between the confinement scales between the two sectors. 
                            This difference can be approximately determined by calculating the value of the dark sector's $\Lambda$ after running upward in energy from $\Lambda_V$ to the lowest breaking scale $M_I$ and then to the second, $M_X$, before evolving 
                            back down in energy until we reach the confinement regime. Using this it can be calculated that at one loop the ratio of the confinement scales is given by
                            \begin{equation}
                             \frac{\Lambda_D}{\Lambda_V} = \frac{M_X}{M_I}^{\frac{b_D-b_V}{b_0}},
                            \end{equation}
                            where the beta functions here are for the intermediate gauge groups in the intermediate range $M_I \le M \le M_X$ for the two sectors, and $b_0$ is the SU(3) beta function given in Eq. 3. This calculation allows us to see that similar but different confinement scales 
                            can be generated from a model with different gauge symmetries at high energy, and for this reason we wish to consider the full set of possible symmetry breaking scenarios. In the next section we will examine what breaking chains are possible in
                            each sector.

                            \section{\bf Multi-Step Breaking Chains}\label{sec:Breakchains}
                            We wish to systematically explore all the possibilities for the different breaking chains that can occur in each sector for an SO(10) model in order to examine which chains 
                            allow for realistic models of both sectors. There are a number of paths through which SO(10) can break down to a gauge theory containing the SM with two of the most notable being through the Pati-Salam $SU(4) \times SU(2) \times SU(2)$\cite{Pati:1973rp} and the Georgi-Glashow SU(5) \cite{Georgi:1974sy} subgroups. 
                            For the visible sector we are mostly concerned with these particular models, however for the dark sector we are free to choose any breaking which leaves unbroken an SU(3) theory at low energy. 
                            This opens up a large number of choices of Higgs multiplet representations in the dark sector. We will limit ourselves to the cases of one and two intermediate scales as 
                            additional scales add complexity without necessarily offering more insight into possible outcomes. Below we list all of the possible breaking chains we can consider for the color force in the dark sector.
                            We consider first of all chains with just one intermediate scale, $M_I$, between the confinement scale, $\Lambda_D$, and the GUT scale $M_X$. These are
                                \begin{equation}
                                \begin{split}
                               SO(10) \rightarrow SO(9) \rightarrow SU(3) \;\;  (I)\\
                               SO(10) \rightarrow SO(8) \rightarrow SU(3)\;\;  (II)\\
                               SO(10) \rightarrow SO(7) \rightarrow SU(3)\;\;  (III)\\
                               SO(10) \rightarrow SU(5) \rightarrow SU(3)\;\;  (IV)\\
                               SO(10) \rightarrow SU(4) \rightarrow SU(3)\;\;  (V)\\ 
                               \end{split}
                               \end{equation}
                            and secondly we consider models with two intermediate scales, $M_I$ and $M_J$, between $M_X$ and the low energy theory, with $M_J \ge M_I$. These are
                               \begin{equation}
                               \begin{split}
                               SO(10) \rightarrow SO(9) \rightarrow SO(8) \rightarrow  SU(3)\;\;  (VI)\\
                               SO(10) \rightarrow SO(9) \rightarrow SO(7) \rightarrow  SU(3)\;\;  (VII)\\
                               SO(10) \rightarrow SO(9) \rightarrow SU(4) \rightarrow  SU(3)\;\;  (VIII)\\
                               SO(10) \rightarrow SO(8) \rightarrow SO(7) \rightarrow  SU(3)\;\;  (IX)\\
                               SO(10) \rightarrow SO(8) \rightarrow SU(4) \rightarrow  SU(3)\;\;  (X)\\
                               SO(10) \rightarrow SO(7) \rightarrow SU(4) \rightarrow  SU(3)\;\;  (XI)\\
                               SO(10) \rightarrow SU(5) \rightarrow SU(4) \rightarrow  SU(3)\;\;  (XII).\\
                                        \end{split}
                                 \end{equation}
                               The chains we consider in the visible sector are most often IV and V as well as the case where the two intermediate scales are close enough that the symmetry breaking effectively happens at one scale, as per
                               $SO(10) \rightarrow SU(3)$. We consider this variety as the limiting case for the magnitude of the difference between the two groups one loop beta functions and is useful for cases where the symmetry breaking chains of the two sectors are in fact
                               the same except for the scales at which breaking occurs. This can be seen as delayed symmetry breaking where at one or more of the scales, $M_X, M_I$ and $M_J$ one sector breaks to a subgroup but the other does not. 
                               The analysis is no different than other examples, it is simply that we contrast some intermediate gauge group's running with that of, for instance, the group $SO(10)$ itself.  
                               In examining results we choose a breaking chain for each sector from the list, but we will limit ourselves to only those choices for which the dark scale runs faster in the intermediate range of the running for the case of one intermediate scale.
                               These cases demonstrate the key aspect of these theories, that the 
                               gauge group of the intermediate energy scale can change the final scale of dimensional transmutation in two SU(3) theories that originate from an originally $\Bbb{Z}_2$ symmetric $G \times G$ theory.
                              
                               For the sake of proton decay limits the intermediate scale of the visible sector $M_I$ must be above experimental constraints. Additionally it is important for 
                               consideration of gauge coupling constant unification in the visible sector which we will return to in Section \ref{sec:Pheno}.
                               The scale at which the dark sector becomes SU(3) is not so constrained, 
                               however if it is significantly lower the confinement scales will distance themselves beyond the desired amount. It may also have consequences for the stability of dark matter depending on other features of the hidden sector.
                               It is also natural to consider the models mentioned where Higgs multiplets that gain the same VEV in each sector allow for the lower intermediate scale to be the same in the two sectors. Beyond this the 
                               next highest intermediate scale $M_J$ is constrained only from above in that $M_J < M_X < M_{Planck}$. In the next section we present a proof that for non-SUSY models
                               asymmetric symmetry breaking can be realized in potentials that give minima which describe any of the model types we discussed above.
                        
                           \section{ \bf Multi-Step Asymmetric Symmetry Breaking}\label{sec:ASB}
                           We will now outline how a Higgs sector can accommodate a large variety of symmetry breaking chains in a GUT model of two sectors. As in \cite{Lonsdale:2014wwa} asymmetric symmetry 
                           breaking can induce non-zero VEVs in Higgs multiplets which have $\Bbb{Z}_2$ partners in the opposing sector that retain a VEV of zero. The simplest example has just two pairs of scalar singlet fields that transform under the $\Bbb{Z}_2$ symmetry as
                           \begin{equation}
                           \phi_1 \leftrightarrow \phi_2 ,   \qquad   \chi_1   \leftrightarrow \chi_2.
                           \end{equation}
                           We can then write down the general potential without loss of generality as 
                           \begin{equation}
                           \begin{split}  
                            V= \lambda_{\phi}({\phi_V}^2 + {\phi_D}^2 - v_{\phi}^2)^2 + \\
                             \kappa_{\phi} ({\phi_V}^2 {\phi_D}^2) + \\    
                             \lambda_{\chi}({\chi_V}^2 + {\chi_D}^2 - v_{\chi}^2)^2 + \\
                             \kappa_{\chi} ({\chi_V}^2 {\chi_D}^2) + \\ 
                             \sigma({\phi_V}^2{\chi_V}^2 + {\phi_D}^2{\chi_D}^2)+ 
                              \rho({\phi_V}^2 + {\chi_V}^2 + {\phi_D}^2 + {\chi_D}^2 - v_{\phi}^2- v_{\chi}^2 )^2,  \\
                           \end{split}
                           \end{equation}
                           where cubic terms are taken to be absent by additional discrete symmetries. If all of the parameters in Eq. 11 are positive then each term in the potential is positive definite and thus minimized if it is equal to zero. The total potential is then minimized by VEVs that break the $\Bbb{Z}_2$ symmetry in such a way that
                           \begin{eqnarray}
                           \braket{\phi_1} =  v_\phi ,   \qquad  \braket{\chi_1} =  0, \nonumber\\           
                            \braket{\phi_2} =  0,    \qquad \braket{\chi_2} =  v_\chi.
                           \end{eqnarray}
                           This minimum is also degenerate with its $\Bbb{Z}_2$ partner where it is $\phi_2$ and $\chi_1$ that gain non-zero VEVs.
                           We can then extend this idea to larger representations of gauge groups by replacing the singlet fields with Higgs multiplets. 
                           The set of Higgs multiplets responsible for symmetry 
                           breaking in each sector can thus be entirely independent for an arbitrary number of representations we add to the theory.
                           Let us firstly take the case of a set of $2n$ singlet scalar fields, ${H_V}_1,{H_D}_1,...,{H_V}_n,{H_D}_n$, where under the $\Bbb{Z}_2$ symmetry,
                           \begin{equation}
                           H_V \leftrightarrow H_D  .
                           \end{equation}             
                           We then consider general potentials where again all of the parameters are positive and each individual term is positive definite and cubic terms are taken to be absent by discrete symmetries. For the case of $n=3$ we have
                           \begin{equation} 
\begin{split}  
 V= \lambda_{H_1}({H_V}_1^2 + {H_D}_1^2 - v_{H_1}^2)^2 + \\
    \kappa_{H_1} ({H_V}_1^2 {H_D}_1^2) + \\    
    \lambda_{H_2}({H_V}_2^2 + {H_D}_2^2 - v_{H_2}^2)^2 + \\
    \kappa_{H_2} ({H_V}_2^2 {H_D}_2^2) + \\ 
    \sigma_1({H_V}_1^2{H_V}_2^2 + {H_D}_1^2{H_D}_2^2)+ 
    \rho_1({H_V}_1^2 + {H_V}_2^2 + {H_D}_1^2 + {H_D}_2^2 - v_{H_1}^2- v_{H_2}^2 )^2 + \\
   \lambda_{H_3}({H_V}_3^2 + {H_D}_3^2 - v_{H_3}^2)^2 + \\
    \kappa_{H_3} ({H_V}_3^2 {H_D}_3^2) + \\ 
    \sigma_3({H_V}_1^2{H_D}_3^2 + {H_D}_1^2{H_V}_3^2)+ 
    \rho_3({H_V}_1^2 + {H_V}_3^2 + {H_D}_1^2 + {H_D}_3^2 - v_{H_1}^2- v_{H_3}^2 )^2 + \\
     \sigma_2({H_V}_3^2{H_V}_2^2 + {H_D}_3^2 {H_D}_2^2)+ 
    \rho_2({H_V}_3^2 + {H_V}_2^2 + {H_D}_3^2 + {H_D}_2^2 - v_{H_3}^2- v_{H_2}^2 )^2.
 \end{split}
\end{equation}
In this case the minimum is given by 
\begin{eqnarray}
 \braket{{H_V}_1} =  v_{H_1} ,   \qquad  \braket{{H_D}_1} =  0,     \nonumber \\            
 \braket{{H_V}_2} =  0,    \qquad \braket{{H_D}_2} =  v_{H_2} ,\\
 \braket{{H_V}_3} =  v_{H_3} ,   \qquad \braket{{H_D}_3} =  0.   \nonumber
\end{eqnarray}
The above minima could have been the reverse where the V and D subscripts are interchanged of course. We simply label the sector which develops the features of the SM as the visible sector.
This potential demonstrates the general procedure by which we can generate non-supersymmetric asymmetric symmetry breaking multi step chains. The first two sets of fields form an asymmetric set as in Eq. 11 and
for any additional field, such as $H_3$ we can choose for it to align with either the visible or dark sector based on these choices:
For coupling between fields that we want to break similarly we set $\sigma$ to couple fields in opposing sectors and the $\rho$ term to be that which allows for same sector terms. In this case we choose for $H_3$ to break the same as $H_1$ so $\sigma_3$ couples fields of different sectors.
Then for mixing between $H_3$ and fields that break differently we
set $\sigma$ to couple the same sector fields, where in Eq. 14 we have $\sigma_2$ coupling same sector fields since $H_2$ is aligned with the opposite sector to $H_1$. Following this simple prescription allows us to add an arbitrary number of multiplets to each sector with the asymmetry determining which sectors will gain the symmetry breaking aspects of that multiplet.
We can then consider representations of SO(10) where now each ${H_V}_n  \sim (R_n, 1) $ and its $\Bbb{Z}_2$ partner transforms as ${H_D}_n \sim (1, R_n)$. The general potential will contain additional couplings, however it will 
always contain an analogous set of terms to those above for which we can always generate an asymmetric array of VEVs. These will then drive the symmetry breaking of the two sectors to be completely different.

As we mentioned earlier the simplest variety of SO(10) model is one where the asymmetry in the VEVs of the potential is not limited to distinguishing between zero and non-zero, but rather creates an asymmetry in 
the size of the VEVs which are all non-zero.
Consider a potential of just two pairs as in Eq. 11 but with each of $\kappa_{\phi}, \kappa_{\chi} < 0$. In this scenario we can create asymmetries of the form 
\begin{eqnarray}
 \braket{\phi_1} =  \braket{\chi_2},   \qquad  \braket{\chi_1} =  \braket{\phi_2}.
\end{eqnarray}
We found it possible to generate a ratio of  $\braket{\phi_1}/\braket{\chi_1} \approx 10^3$ for a very constrained region of parameter space. 
Such a potential can minimally accommodate exactly the number of Higgs multiplets necessary to break two copies of SO(10) to the same final gauge group but with different gauge 
groups in the intermediate range depending on the choice of Higgs multiplet.
While simple in the number of multiplets, this minimal theory suffers from a much smaller allowed parameter space than the previously discussed ASB mechanisms. In particular the size of parameters must be fine tuned slightly such that we very nearly have $\kappa_\chi \simeq \kappa_\phi$ and $-\kappa_\phi -\kappa_\chi 
\simeq \sigma$. If we remove the condition of having just two breaking scales and allow each of the four fields to attain different VEVs then a much broader range of the parameter space is compatible.

We can also develop models in which additional pairs of multiplets that transform under the $\Bbb{Z}_2$ symmetry are added as in Eq. 14 but break in such a way that they both gain non-zero VEVs and thus both contribute to the symmetry breaking in each sector.
This is in fact the simplest method in some kinds of cases where the same dimensional representation is useful for the symmetry breaking needed in each sector.
This can be  always be accomplished by, for example, having these added fields couple only weakly to the previously added fields.
\\\\
We illustrate this asymmetric breaking with a particular $SO(10) \times SO(10)$ potential which breaks the mirror symmetric GUT group to $[SU(4) \times SU(2) \times SU(2)]_V \times [SU(5) \times U(1)]_D$. 
Within the context of the standard model, such a theory would need at least one more Higgs multiplet in order to break the Pati-Salam group to $SU(3) \times SU(2) \times U(1)$. Within our variety of 
models we would require at least one additional mirror symmetric pair of representations to break the symmetry in each sector to one containing SU(3). 
Since the important results from this work are the generation of different symmetries in the intermediate range we focus on constructing a potential that asymmetrically generates the first step of the breaking chain.
We consider a set of fields transforming as
\begin{eqnarray}
\phi_V \sim (45,1),  \qquad  \chi_V \sim (54,1), \nonumber\\ 
\phi_D \sim (1,45),  \qquad  \chi_D \sim (1,54).
\end{eqnarray}
With these we can follow the procedure detailed in the toy model and construct an asymmetric potential. 
Each of the terms in the toy model has a direct analogue and in addition to these there will be new terms from unique contractions of the Higgs multiplets. The general renormalizable fourth order potential is
\begin{eqnarray}
-\frac{\mu_{\phi}^2}{2} ({\phi_V}_{ij}{\phi_V}_{ji} +{\phi_D}_{ij}{\phi_D}_{ji}) + \frac{\lambda_{\phi}}{4} (({\phi_V}_{ij}{\phi_V}_{ji})^2 +({\phi_D}_{ij}{\phi_D}_{ji})^2) + \nonumber \\
\frac{\alpha_{\phi}}{4}({\phi_V}_{ij}{\phi_V}_{jk}{\phi_V}_{kl}{\phi_V}_{li} + {\phi_D}_{ij}{\phi_D}_{jk}{\phi_D}_{kl}{\phi_D}_{li})+ \kappa_{\phi}({\phi_D}_{ij}{\phi_D}_{ji}{\phi_V}_{kl}{\phi_V}_{lk}) + \nonumber \\
-\frac{\mu_{\chi}^2}{2} ({\chi_V}_{ij}{\chi_V}_{ji} +{\chi_D}_{ij}{\chi_D}_{ji}) +\frac{\lambda_{\chi}}{4} (({\chi_V}_{ij}{\chi_V}_{ji})^2 +({\chi_D}_{ij}{\chi_D}_{ji})^2 ) + \nonumber \\
\frac{\alpha_{\chi}}{4}({\chi_V}_{ij}{\chi_V}_{jk}{\chi_V}_{kl}{\chi_V}_{li} + {\chi_D}_{ij}{\chi_D}_{jk}{\chi_D}_{kl}{\chi_D}_{li}) +  \kappa_{\chi}({\chi_D}_{ij}{\chi_D}_{ji}{\chi_V}_{kl}{\chi_V}_{lk}) + \nonumber \\
\frac{\beta \mu_{\chi}}{3}({\chi_V}_{ij}{\chi_V}_{jk}{\chi_V}_{ki} + {\chi_D}_{ij}{\chi_D}_{jk}{\chi_D}_{ki})+ 
c_1({\phi_D}_{ij}{\phi_D}_{ji}{\chi_V}_{kl}{\chi_V}_{lk} +{\phi_V}_{ij}{\phi_V}_{ji}{\chi_D}_{kl}{\chi_D}_{lk}  )+ \nonumber \\
c_2({\phi_D}_{ij}{\phi_D}_{ji}{\chi_D}_{kl}{\chi_D}_{lk} +{\phi_V}_{ij}{\phi_V}_{ji}{\chi_V}_{kl}{\chi_V}_{lk}  )+ 
c_3({\phi_D}_{ij}{\phi_D}_{jk}{\chi_D}_{kl}{\chi_D}_{li} +{\phi_V}_{ij}{\phi_V}_{jk}{\chi_V}_{kl}{\chi_V}_{li}  )+ \nonumber \\
c_4({\phi_D}_{ij}{\phi_D}_{jk}{\chi_D}_{ki}) +{\phi_V}_{ij}{\phi_V}_{jk}{\chi_V}_{ki}  ) + 
c_5( Tr[({\phi_V}_{ik}{\chi_V}_{km} - {\chi_V}_{il}{\phi_V}_{lm})^2] + Tr[({\phi_D}_{ik}{\chi_D}_{km} - {\chi_D}_{il}{\phi_D}_{lm})^2] ). \nonumber \\
\end{eqnarray}
The addition of the cubic term is necessary for the pattern of symmetry breaking we have chosen. This differs from  the toy model cases where an additional $\Bbb{Z}_2$ symmetry protected the potentials from such cubic terms. 
Relaxing this condition still allows for asymmetric solutions for the VEVs of the two sectors however as discussed in appendix A. For the sake of simplicity we also set the parameters $c_3,c_4,c_5$ to be zero as 
large values will remove the asymmetric VEV structure. 
The analysis can be simplified by transforming the fields into a simplified VEV form. For the adjoint representation this becomes a block 
diagonal matrix with each block being a $2 \times 2$ antisymmetric matrix. For the 54 we have a traceless diagonal matrix. 
For the region of parameter space discussed in the appendix the potential is minimized with VEVs
\begin{eqnarray}
\braket{\phi_V} & = & M_i
\begin{pmatrix}
0 & a & 0 & 0 & 0 & 0 & 0 & 0 & 0 & 0 \\
-a & 0 & 0 & 0 & 0 & 0 & 0 & 0 & 0 & 0 \\
0 & 0 & 0 & a & 0 & 0 & 0 & 0 & 0 & 0 \\
0 & 0 & -a & 0 & 0 & 0 & 0 & 0 & 0 & 0 \\
0 & 0 & 0 & 0 & 0 & a & 0 & 0 & 0 & 0 \\
0 & 0 & 0 & 0 & -a & 0 & 0 & 0 & 0 & 0 \\
0 & 0 & 0 & 0 & 0 & 0 & 0 & a & 0 & 0 \\
0 & 0 & 0 & 0 & 0 & 0 & -a & 0 & 0 & 0 \\
0 & 0 & 0 & 0 & 0 & 0 & 0 & 0 & 0 & a \\
0 & 0 & 0 & 0 & 0 & 0 & 0 & 0 & -a & 0 \\
 \end{pmatrix} \nonumber \\
\braket{\phi_D} & = & 0 \nonumber \\
\braket{\chi_V} & = & 0 \nonumber \\
\braket{\chi_D} & = & M_j
 \begin{pmatrix}
b & 0 & 0 & 0 & 0 & 0 & 0 & 0 & 0 & 0 \\
0 & b & 0 & 0 & 0 & 0 & 0 & 0 & 0 & 0 \\
0 & 0 & b & 0 & 0 & 0 & 0 & 0 & 0 & 0 \\
0 & 0 & 0 & b & 0 & 0 & 0 & 0 & 0 & 0 \\
0 & 0 & 0 & 0 & b & 0 & 0 & 0 & 0 & 0 \\
0 & 0 & 0 & 0 & 0 & b & 0 & 0 & 0 & 0 \\
0 & 0 & 0 & 0 & 0 & 0 & c & 0 & 0 & 0 \\
0 & 0 & 0 & 0 & 0 & 0 & 0 & c & 0 & 0 \\
0 & 0 & 0 & 0 & 0 & 0 & 0 & 0 & c & 0 \\
0 & 0 & 0 & 0 & 0 & 0 & 0 & 0 & 0 & c \\
 \end{pmatrix}.
\end{eqnarray}
In the above $SO(10)_V$ breaks by the VEV of the \textbf{54} to $SO(6) \times SO(4) \sim SU(4) \times SU(2) \times SU(2)$ and the \textbf{45} 
serves to break $SO(10)_D$ to $SU(5) \times U(1)$.\cite{Georgi:1982jb, Wu:1981eb}
Following this symmetry breaking we would then need additional Higgs multiplets to break each of the gauge groups to SU(3) colour theories after which the running couplings will be parallel.
Due to the complexity of analyzing potentials with increasing numbers of large dimensional Higgs multiplets we leave such detailed models to more specific theories.
We have however completed our stated objective of constructing an SO(10) asymmetric potential, built according to the principles of ASB, and showing that by choosing the breaking scales in 
the two sectors and the breaking chains listed in the previous section, asymmetric potentials can be constructed such that exactly that 
scenario is the minimum of the potential.   
In the next section we attempt to generalize such possibilities for supersymmetric models. 
We specifically look at the general case of real representations which we can examine in an illustrative model.
                         
                         \section{ \bf Supersymmetric Theories}\label{sec:SUSY}
                          As in \cite{Lonsdale:2014wwa} this analysis is predicated on the unification of coupling constants and for this, among other reasons such as the gauge hierarchy problem, we will explore 
                          supersymmetric varieties of these models in this section. 
                          As discussed in \cite{Lonsdale:2014wwa}, Supersymmetric ASB requires more fields than the non-SUSY case, specifically gauge singlets. Here we will outline a general scheme to create asymmetric symmetry 
                          breaking chains from the superpotential. In general additional fields are required to allow for the scalar potential to have the necessary terms that drive ASB since only including non-singlet Higgs multiplets does not allow us to couple fields from the different sectors at all in the scalar potential.
                          The method that we outline below is not necessarily the simplest way to generate such breaking for any specific choice of representations or breaking chains; indeed for many simple models as few as one additional singlet is required \cite{Lonsdale:2014wwa}. 
                          The purpose of this discussion is to provide an existence proof that for any symmetry breaking chain we may consider in Section \ref{sec:Results}, a 
                          scalar potential can be created which allows for such a vacuum solution.

                          We wish to consider a supersymmetric extension to the argument of Section \ref{sec:ASB} wherein pairs of Higgs multiplets can be added one at a time to a model in a $\Bbb{Z}_2$ symmetric manner while 
                          allowing us to choose which sector its VEVs will favor by appropriate choice of couplings. 
                          Take the case of the fields ${H_1}_V, {H_1}_D, {H_2}_V, {H_2}_D, {H_3}_V, {H_3}_D, X_1, X_2, Y_1, Y_2, Z_1, Z_2, \phi, \theta,$ where under the $\Bbb{Z}_2$ symmetry
                         \begin{eqnarray}
                          X_1 \leftrightarrow X_2 ,    \qquad   Y_1   \leftrightarrow Y_2, \nonumber \\
                          Z_1 \leftrightarrow Z_2 ,    \qquad  \phi   \leftrightarrow \phi , \qquad  \theta   \leftrightarrow \theta.
                         \end{eqnarray}
                          We then consider the general, renormalizable, gauge invariant superpotential that respects the $\Bbb{Z}_2$ symmetry between the sectors. In this case we are assuming that the Higgs multiplets form real representations though a similar argument likely exists for complex representations as well. 
                          We do not write down all of the terms in such a superpotential, only those which directly contribute to the asymmetric symmetry breaking terms as in Eq. 14:
                          \begin{eqnarray}
                                 W & = & \rho_{1}({H_2}_V^2 \phi + {H_2}_D^2 \phi) + 
                                  \rho_{2} ({H_2}_V^2 Y_1 + {H_2}_D^2 Y_2) + 
                                  \rho_{3}({H_1}_V^2 Z_2 + {H_1}_D^2 Z_1) +
                                  \rho_{4}({H_2}_V^2 Z_2 + {H_2}_D^2 Z_1)  \nonumber \\
                                & + &  \rho_{5}({H_1}_V^2 \theta + {H_1}_D^2 \theta) +
                                   \rho_{6} ({H_1}_V^2 X_1 + {H_1}_D^2 X_2) + 
                                  \rho_{7}( {Z_1}^3 + {Z_2}^3)   + 
                                 \rho_{8} {\theta}^3 + \rho_{9} {\phi}^3  \nonumber \\
                                & + & \rho_{10}( {X_1}^3 + {X_2}^3)  +  
                                 \rho_{11}( {Y_1}^3 + {Y_2}^3) +  
                                   \rho_{12}({H_3}_V^2 X_2 + {H_3}_D^2 X_1) + 
                                  \rho_{13}({H_3}_V^2 Y_1 + {H_3}_D^2 Y_2) \nonumber \\
                                & + & \rho_{14}({H_3}_V^2 \theta + {H_3}_D^2 \theta) +
                                  \rho_{15}({H_3}_V^2 \phi + {H_3}_D^2 \phi) + ...
                              %
                            %
                           %
                            %
                           %
                          %
                          %
                           %
                           %
                            %
                           %
                             %
                             %
                          \end{eqnarray}
                           The scalar potential then comes from the sum of soft terms and ${W^{i}}^{*} W_i$ where we ignore the D-terms for this analysis, though in general such terms will add positive definite quartic interactions 
                         among those fields which are non-singlets which will not negatively affect the results discussed here.
                         We examine the extreme case of the parameter space where the terms shown dominate and all other parameters in the superpotential are at or very close to zero.
                         In this case the scalar potential minimally contains only those terms that 
                         would exist without the purely singlet fields, as in Eq. 14 and which are necessary for ASB, in addition to a number of other terms which contain the purely singlet fields. If the sum of 
                         the soft mass terms and mass terms from the superpotential F-terms for the singlet fields $X, Y, Z, \theta, \phi$ is positive then these fields can maintain a VEV of zero at the minimum. 
                         In this case the dependencies among the remaining fields is entirely that of N pairs of fields under the $\Bbb{Z}_2$ symmetry exactly like that of Section \ref{sec:ASB} where the symmetry 
                         breaking of added fields can be chosen by the couplings to the previously added fields and we only have quartic and quadratic terms to deal with. Again we have that the symmetry breaking of an added multiplet such as $H_3$ can be chosen by its coupling strength to previously added fields, in this case the $X$ and $Y$ fields. In Eq. 21 we have chosen to include
                         the couplings for which, upon taking the derivatives with respect to $X_1, Y_1, X_2, Y_2$ create the terms that follow the prescription discussed in Section \ref{sec:ASB}. If one wished to have the field $H_3$ break similarly to $H_2$ instead of $H_1$ we simply reduce the magnitude of
                         the parameters $\rho_{12, 13}$ and replace them with larger couplings for the terms $({H_3}_V^2 X_1 + {H_3}_D^2 X_2)$ and $({H_3}_V^2 Y_2 + {H_3}_D^2 Y_1)$ which were previously among the omitted terms.
                         $Z_1$ and $Z_2$ set the initial asymmetry between the first two pairs of fields ${H_1}_{V,D}$ and ${H_2}_{V,D}$ while the 
                         F-terms from $\theta$ and $\phi$ create the remaining couplings in the $\rho$ terms from Eq. 14.
                         The magnitude of the VEVs of the Higgs multiplets will however depend on the size of the soft mass terms that we add and so it may be difficult to construct models with very different mass scales. 
                         This  may however work to our benefit as large differences in the values of $\Lambda_{QCD}$ can be generated in short ranges if the difference in the beta functions is large. One can take this example as a proof of concept that asymmetric 
                         models of any number of Higgs multiplets can be built in SUSY with the addition of singlet fields.
                         Now that we have demonstrated such possible models in both supersymmetric and non supersymmetric cases we will move on to displaying the numerical results for the dark confinement scale for different choices of 
                         representations of the Higgs multiplets.

                          \section{Dark QCD scale from asymmetric symmetry breaking}\label{sec:Results}
                          We firstly consider the set of models with just one intermediate scale which allows just one energy range over which the beta functions of the two SU(3) groups differ. In this case we are 
                          thus only considering models where the group in the dark sector has a larger beta function. We consider both SUSY and non-SUSY models here since for this part of the analysis the only discerning feature is the size of the beta functions which for the SUSY case contains supersymmetric partners to consider as per Eq. 2. 
                          We take the unification point to be where both sectors become $SO(10)$, the GUT scale $M_X$ in our context.
                          We can however have cases where the dark sector remains as an SO(10) for the range between $M_X$ and the intermediate scale $M_I$ while the visible sector changes group. The analysis is the same with the intermediate gauge group of the dark sector being simply SO(10).
                       
                          There are three possibilities for the visible sector's QCD parent group. It can remain SU(3) up until $M_X$ while the dark sector changes at $M_I$ or it can become $SU(4)$ or $SU(5)$ at the 
                          $M_I$ and continue to the unification point. For the dark sector group we examined the cases of the chains from Section \ref{sec:Breakchains}. For the case of just two scales $M_X$ and $M_I$ we plot the ratio of confinement scales by using Eq. 7. 
                          We look at the scale $M_I$ and the difference between the two scales $\delta M = M_X - M_I$. 
                          We display in Figures 2-3 the minimal and maximal cases in terms of group choice, that is the largest and smallest difference in beta functions for each of the possible breaking chains in the VS. The color scale of each graph gives the ratio $\frac{\Lambda_D}{\Lambda_V}$. We see in these figures that quite a large 
                          range in the distance between the breaking scales is acceptable if the beta functions are not very different in size, for example in the case of SU(3) and SU(4). The magnitude of this difference may be smaller depending on the particle content of a specific theory though the $\Bbb{Z}_2$ symmetry between the sectors prevents these matter terms in Eq. 2 from generating large differences. For the limiting case of SU(3) and SO(10), on the other hand, we have a much more 
                          constrained parameter space for the choice of breaking scales. 
          
                         \begin{figure}[!ht]
  \begin{minipage}{\textwidth}
    \centering
    \includegraphics[width=.4\textwidth]{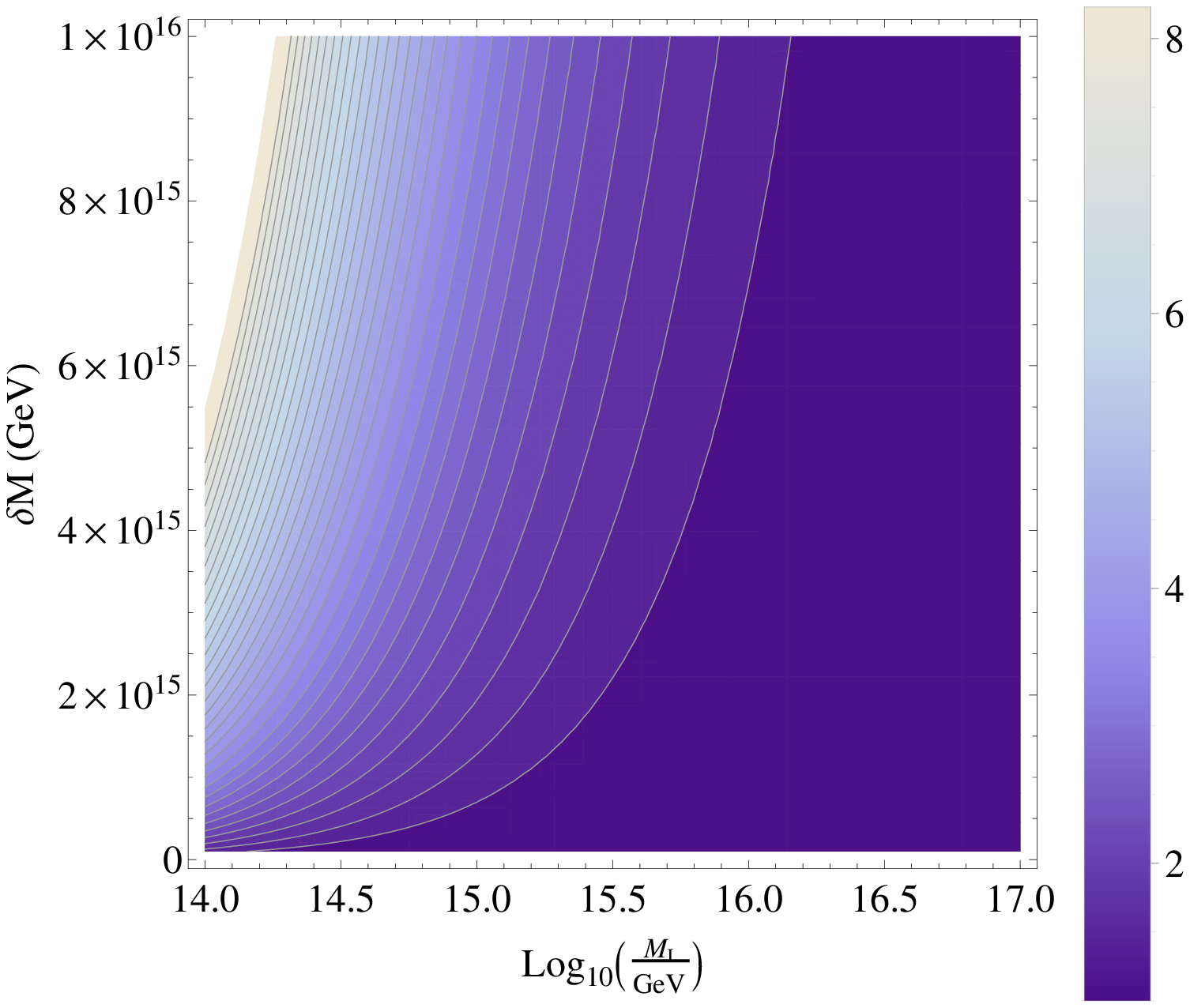}\quad 
    \includegraphics[width=.4\textwidth]{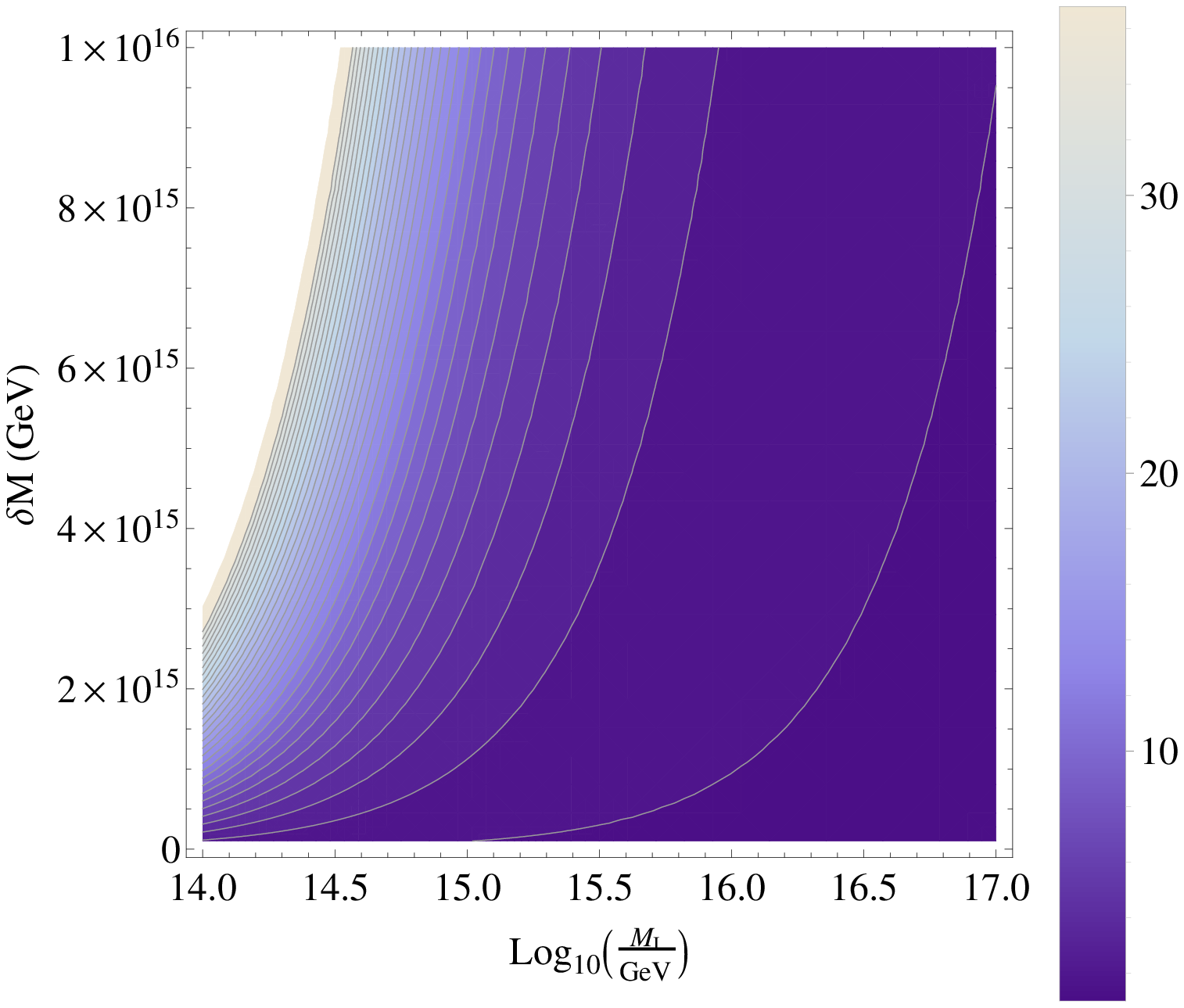}\\   
    \includegraphics[width=.4\textwidth]{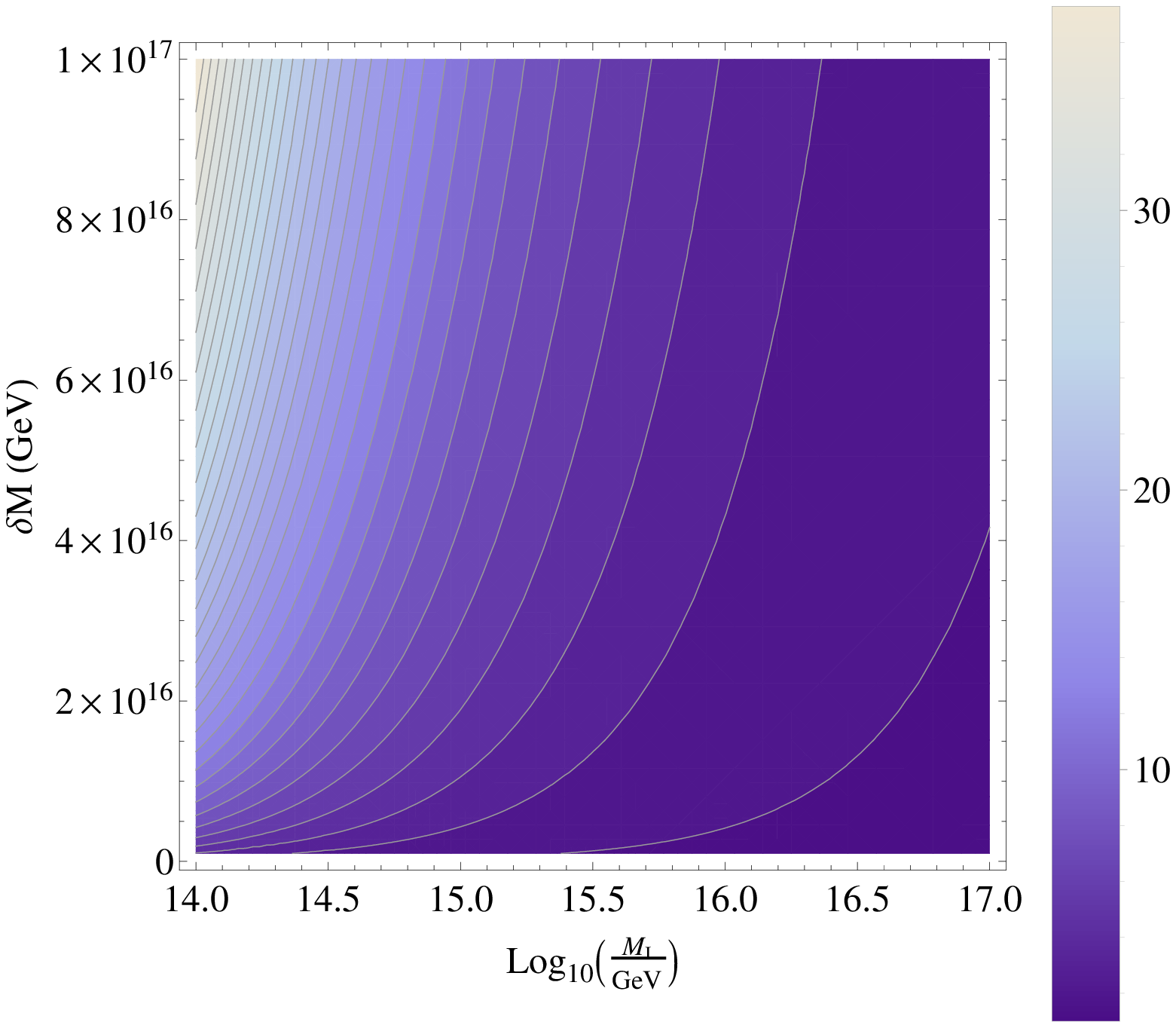}\quad 
    \includegraphics[width=.4\textwidth]{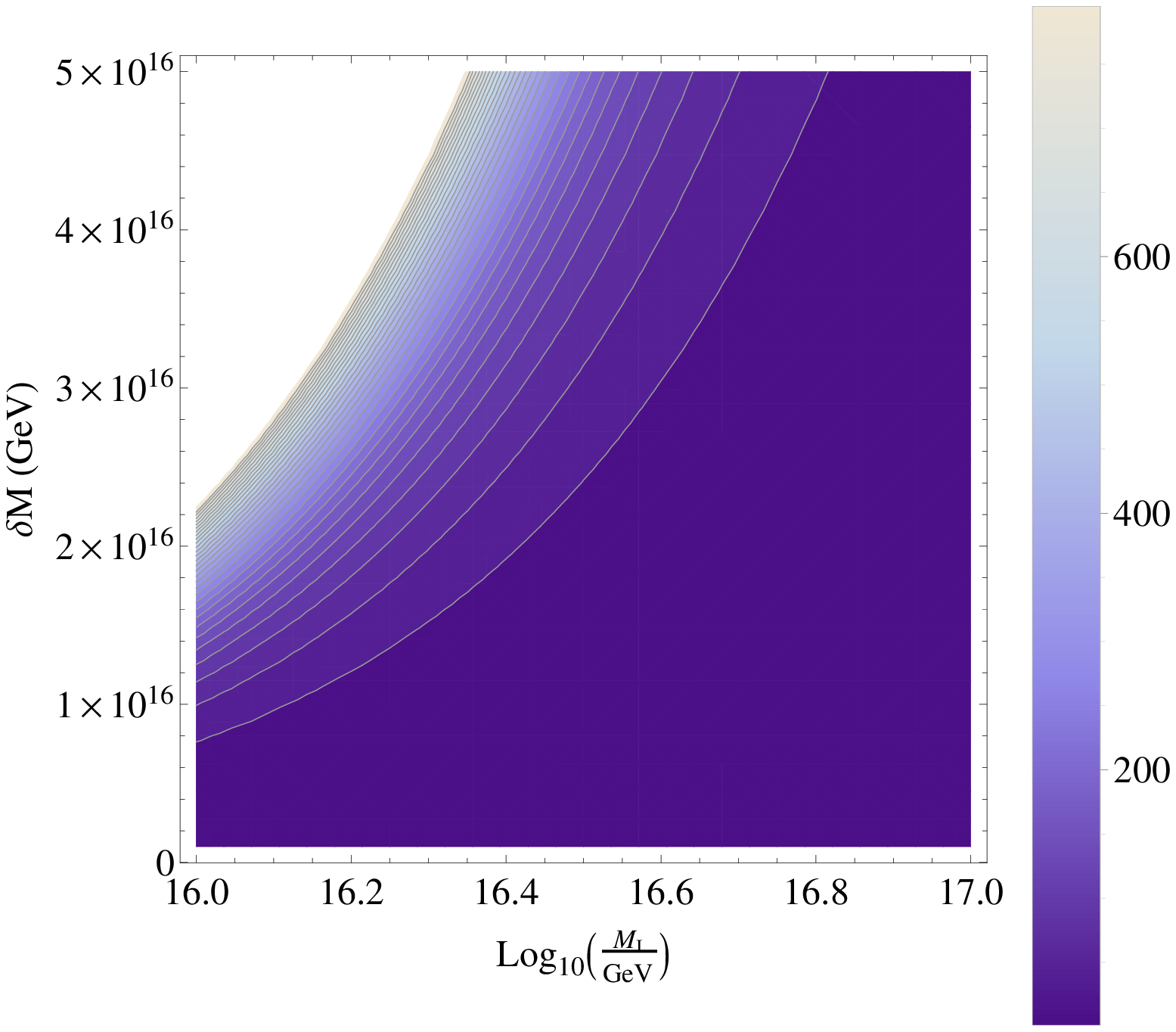}      
    \caption{The ratio of confinement scales  $\Lambda_D/\Lambda_V$ in the two sectors for  a sample of non-supersymmetric breaking chains. The top left figure is generated from $SU(3)_V$ and $SU(4)_D$ as the groups above the scale $M_I$. 
    The top right features $SU(3)_V$ and $SU(5)_D$ above $M_I$ while the bottom left has $SU(4)_V$ and $SU(5)_D$ followed by the bottom right with $SU(3)_V$ and $SO(10)_D$. 
    In each graph the vertical scale is $\delta M$ while the horizontal scale is $M_I$ below which both sectors contain SU(3) subgroups.}
    \label{fig:sub1}
  \end{minipage}\\[1em]
\end{figure}

                         \begin{figure}[!ht]
  \begin{minipage}{\textwidth}
    \centering
    \includegraphics[width=.4\textwidth]{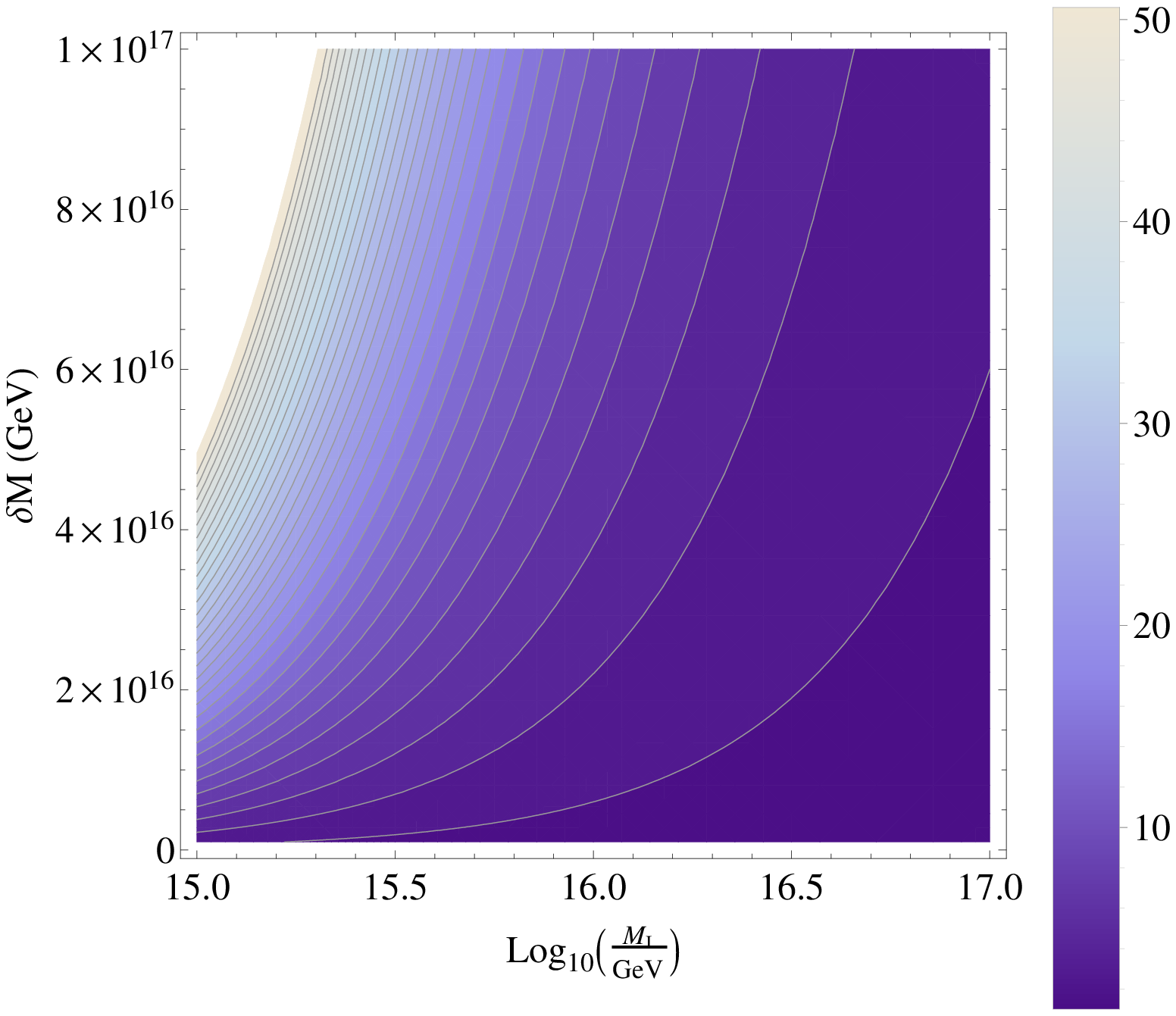}\quad 
    \includegraphics[width=.4\textwidth]{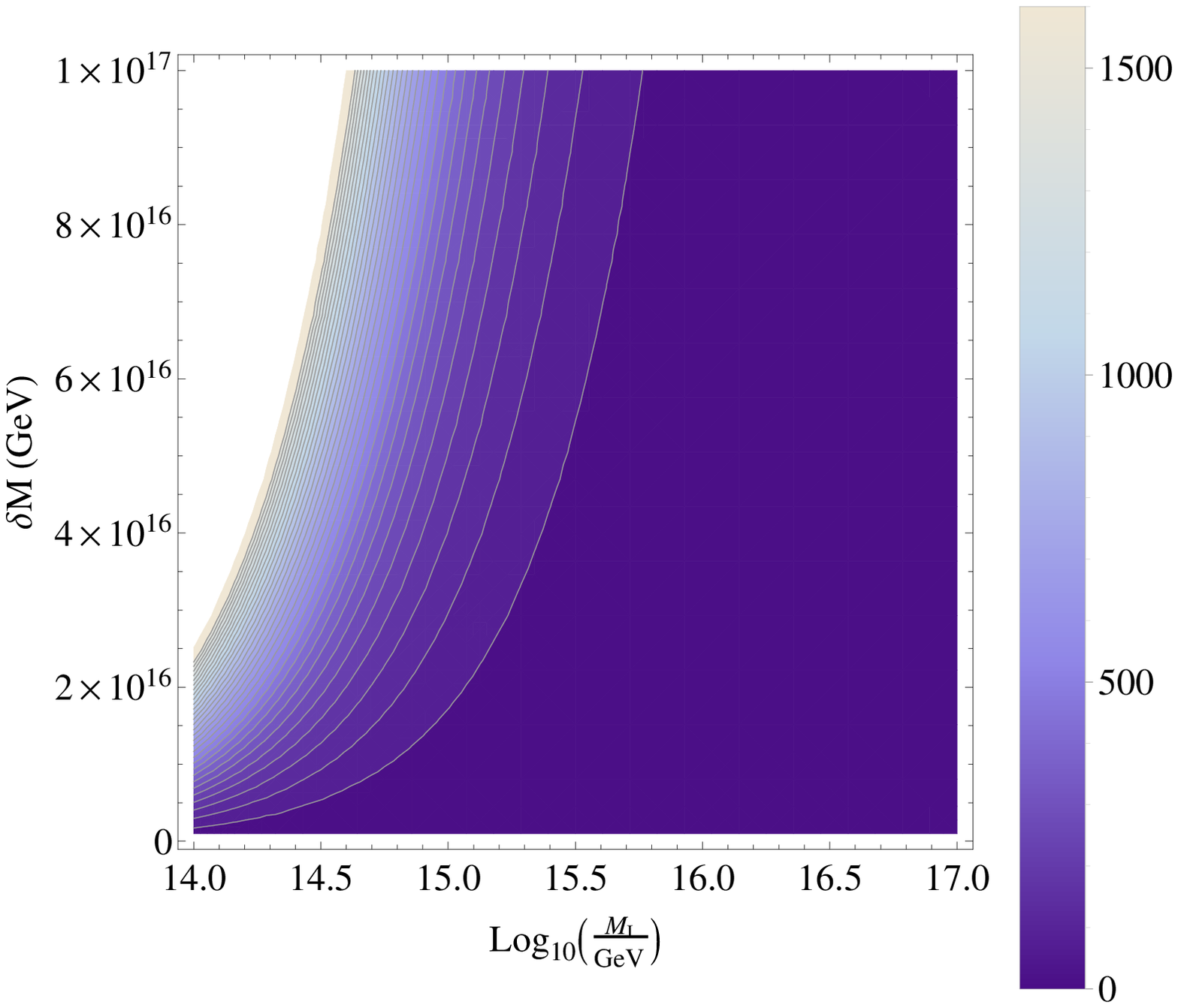}\\   
    \includegraphics[width=.4\textwidth]{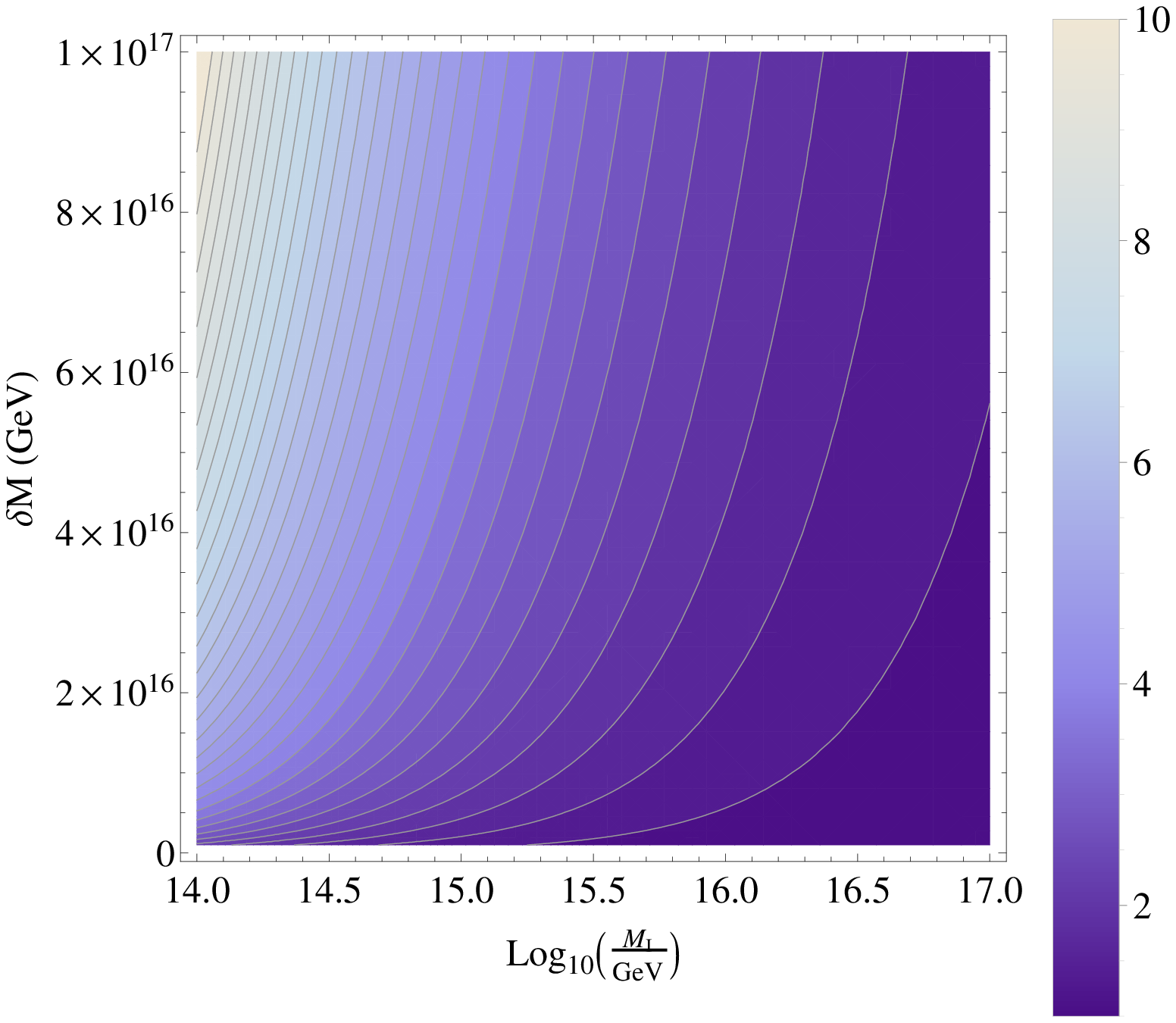}\quad 
    \includegraphics[width=.4\textwidth]{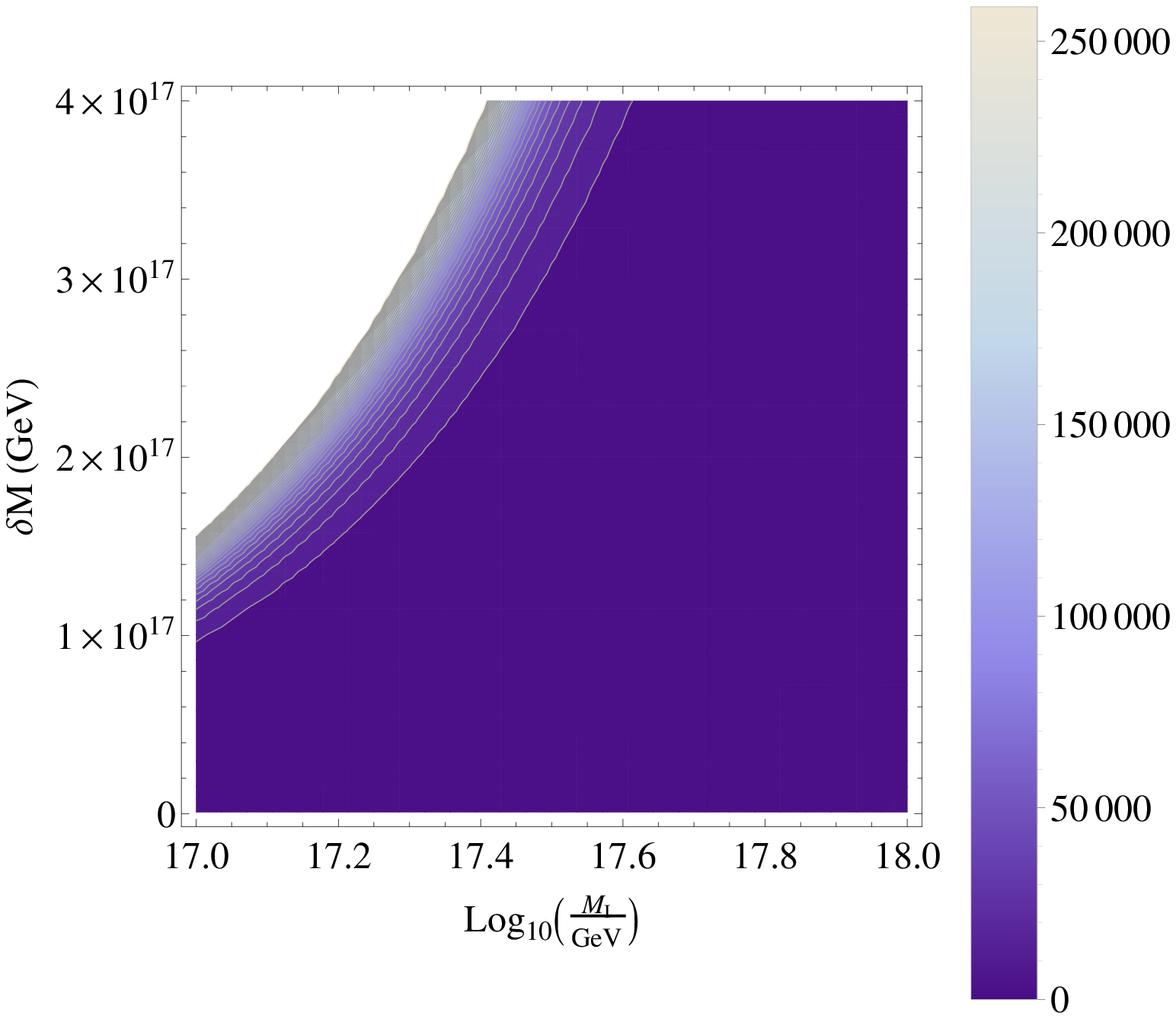}      
    \caption{The ratio of confinement scales  $\Lambda_D/\Lambda_V$ in the two sectors for  a sample of supersymmetric breaking chains. The top left figure is generated from $SU(3)_V$ and $SU(4)_D$ as the groups above the scale $M_I$. 
    The top right features $SU(3)_V$ and $SU(5)_D$ above $M_I$ while the bottom left has $SU(4)_V$ and $SU(5)_D$ followed by the bottom right with $SU(5)_V$ and $SO(10)_D$. }
    \label{fig:sub1}
  \end{minipage}\\[1em]
\end{figure}
                 In these cases the results follow from that of the one intermediate scale case, that is, the final difference in the confinement scales is a function of length of the range over which the couplings run at different rates, 
                 and the magnitude of the difference between the beta functions. Because of this it is possible to create a dark sector with an acceptable confinement scale for any breaking chain that is needed to satisfy visible sector GUT constraints. For example,
                 if a specific model requires a large range between the $SO(10)$ scale $M_X$ and the $SU(5)$ scale $M_I$ in a theory like that of breaking chain IV, then we can choose the scale that the dark sector breaks to SU(5) to be  similar to $M_X$ and run as SU(5) down to a lower scale than $M_I$.
                 We have seen that there are a large number of possible cases for the breaking chains in each sector where the confinement scale in the dark sector is just larger than that of the visible sector. 
                 We have however been treating our GUT scale $M_X$ and intermediate scale$ M_I$ as free parameters and so in the next section we will look to constraining the realistic models and look towards possible future work in this area.           
          
                          \section{Phenomenological Constraints}\label{sec:Pheno}
                          The methods detailed here for generating dark sectors with baryons of a mass scale just above that of the proton are generalizable to many breaking chains and GUT models, not all of which will satisfy phenomenological constraints 
                          such as current proton decay limits. Here we briefly review some of the recent SO(10) GUT models which can satisfy proton decay constraints in the visible sector. Proton decay bounds typically bounds push the scale of unification in SU(5) theories up to energy regimes consistent with the unification of the gauge coupling constants.
                          In some works such as \cite{Dorsner:2005ii} the SU(5) scale is as low as $M_X$ $\approx$ $4 \times 10^{15}$  after the addition of extra Higgs multiplets.
                          In particular we examine some of the recent work on proton decay constraints in GUT models from \cite{Senjanovic:2009}, \cite{Kolesova:2014mfa} where while minimal SU(5) theories are ruled out, 
                          supersymmetric SU(5) theories may still be viable while both supersymmetric and non-supersymmetric SO(10) models can generate cases where proton decay is within experimental limits. We have not gone into any depth on any 
                          specific choice of representations in this paper so it remains an open question how a particular model of ASB can work in the context of these phenomenological constraints.
                           The construction of realistic models also requires the unification of the coupling constants which places strict constraints on the scale at which the visible sector's QCD parent group starts. 
                           We examine such examples for the MSSM running and a non-SUSY case. Below we examine the development of a dark QCD in an extension of this model where the SM gauge couplings unify at an intermediate scale and the two sectors unify closer to the Planck scale.
                           Figure 4 shows the case where we have chain IV in the VS and chain X in the DS. This could be accomplished with \textbf{45} and \textbf{16} or \textbf{126} Higgs multiplets in the VS, together with a  \textbf{54} or \textbf{210'} and \textbf{16} or \textbf{126} in the DS.  
                           Figure 5 shows the direct breaking $SO(10) \rightarrow SU(3)$ for the color force in the VS and chain XII in the DS which was discussed in Section \ref{sec:SO(10)}.

                                                                            \begin{figure}[!ht]
  \begin{minipage}{\textwidth}
    \centering
    \includegraphics[width=.4\textwidth]{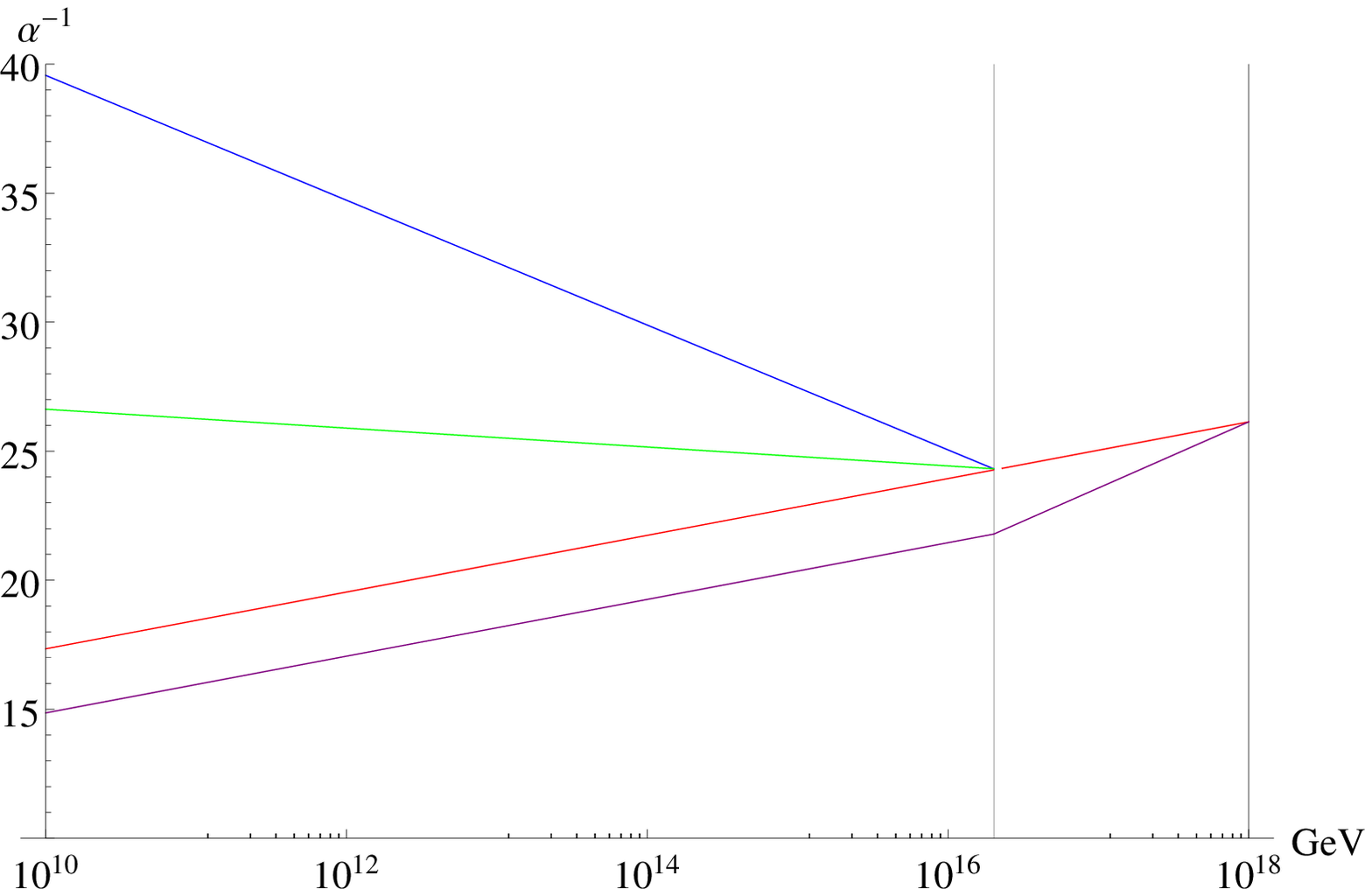}\quad 
    \includegraphics[width=.4\textwidth]{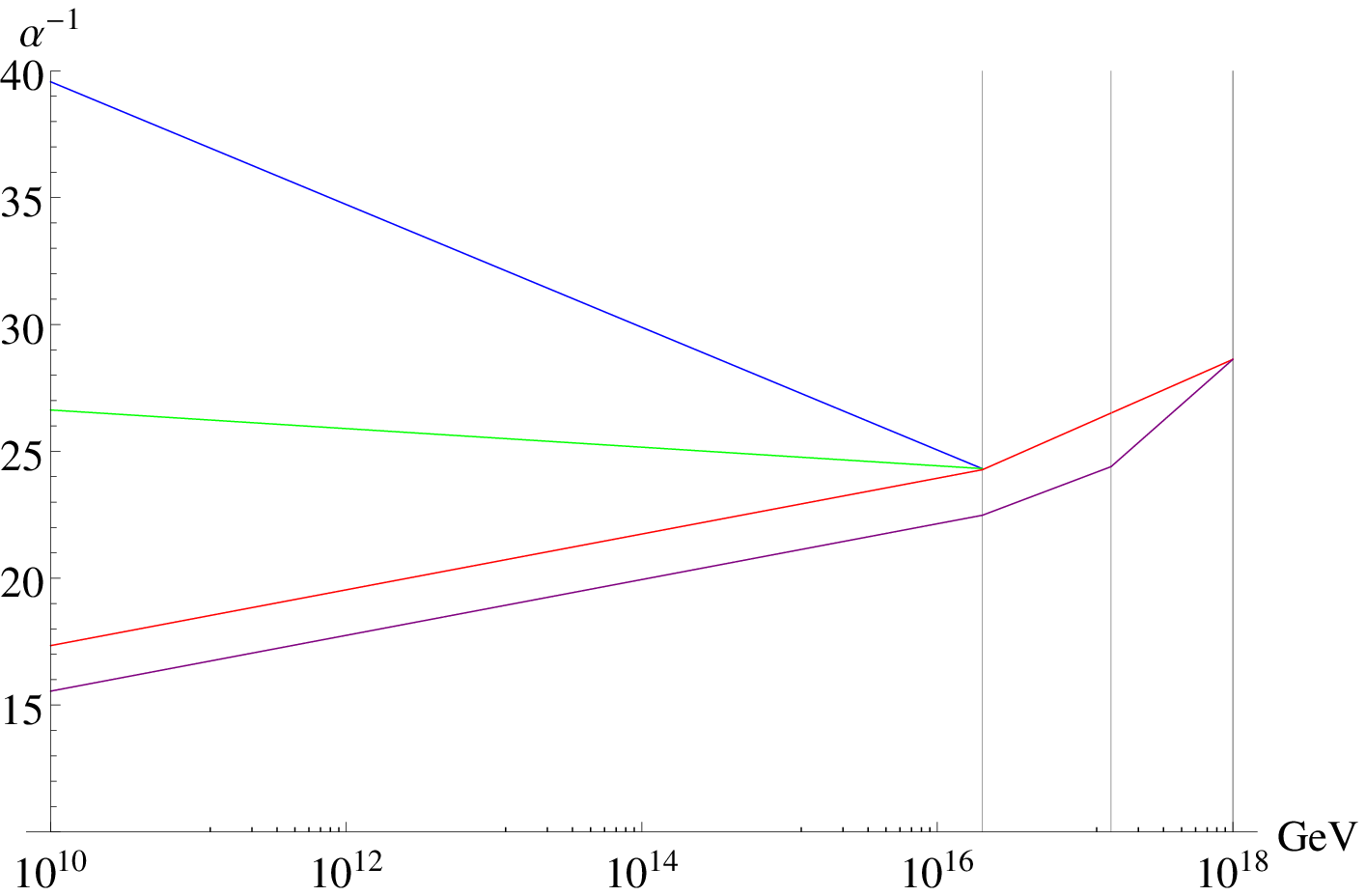}\\   
    \caption{Supersymmetric model with one and two intermediate scales. In the left plot we have $SU(5)$ in the visible sector and $SO(8)$ in the dark above the scale $M_X \approx 10^{16}$ GeV while the right plot shows $SU(5)$ in the visible sector and $SO(8)$ breaking to $SU(4)$ at the scale $M_J \approx 10^{17}$ GeV in the dark sector. Each graph displays the running coupling of the SM forces ($\alpha_1, \alpha_2, \alpha_3$) from top to bottom and that of the color force in the dark sector(the lowest line). 
    The value of the dark confinement scale is 4.1 GeV and 1.9 GeV for the left and right cases, respectively.}
    \label{fig:sub1}
  \end{minipage}\\[1em]
\end{figure}
                          In the MSSM, once we have fixed the scale at which the VS SU(3) is absorbed into SU(5), $M_X$ and any intermediate scale of the dark sector can then be treated as free parameters to generate the dark confinement scale.
                          For the non-SUSY case we examine the work of \cite{Babu:2012vc} in which a non-SUSY SO(10) model with a color sextet allows for the unification of the gauge coupling constants. In this case we can also examine a two step process which 
                          has one segment working to diverge the couplings after SO(10) breaking, while the next part of the breaking regime brings the couplings closer again to result in a dark QCD scale just one order of magnitude greater than the SM 
                          for breaking scales which span over four orders of magnitude.
                
                                                   \begin{figure}[!ht]
  \begin{minipage}{\textwidth}
    \centering
    \includegraphics[width=.4\textwidth]{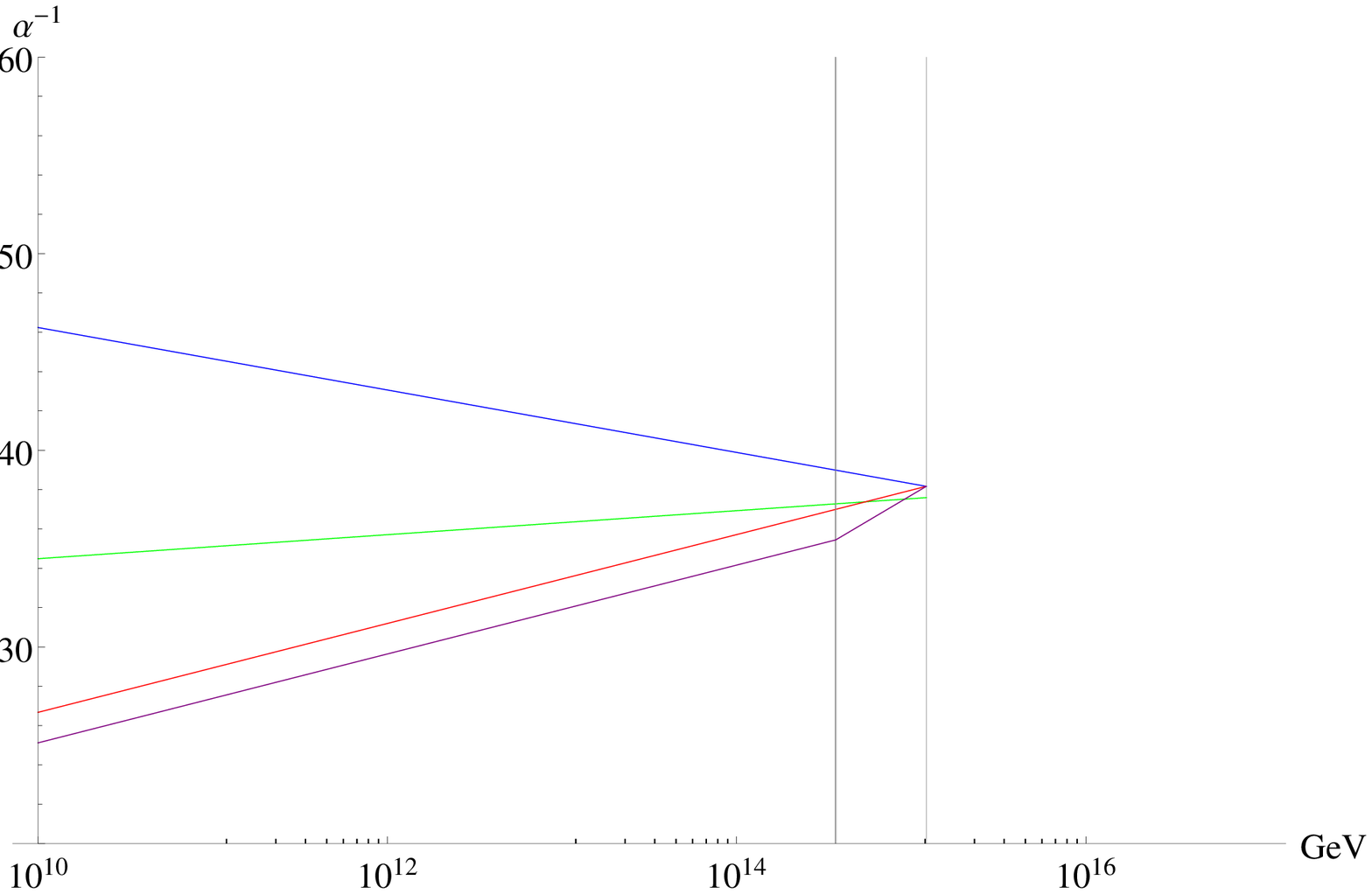}\quad 
  \includegraphics[width=.4\textwidth]{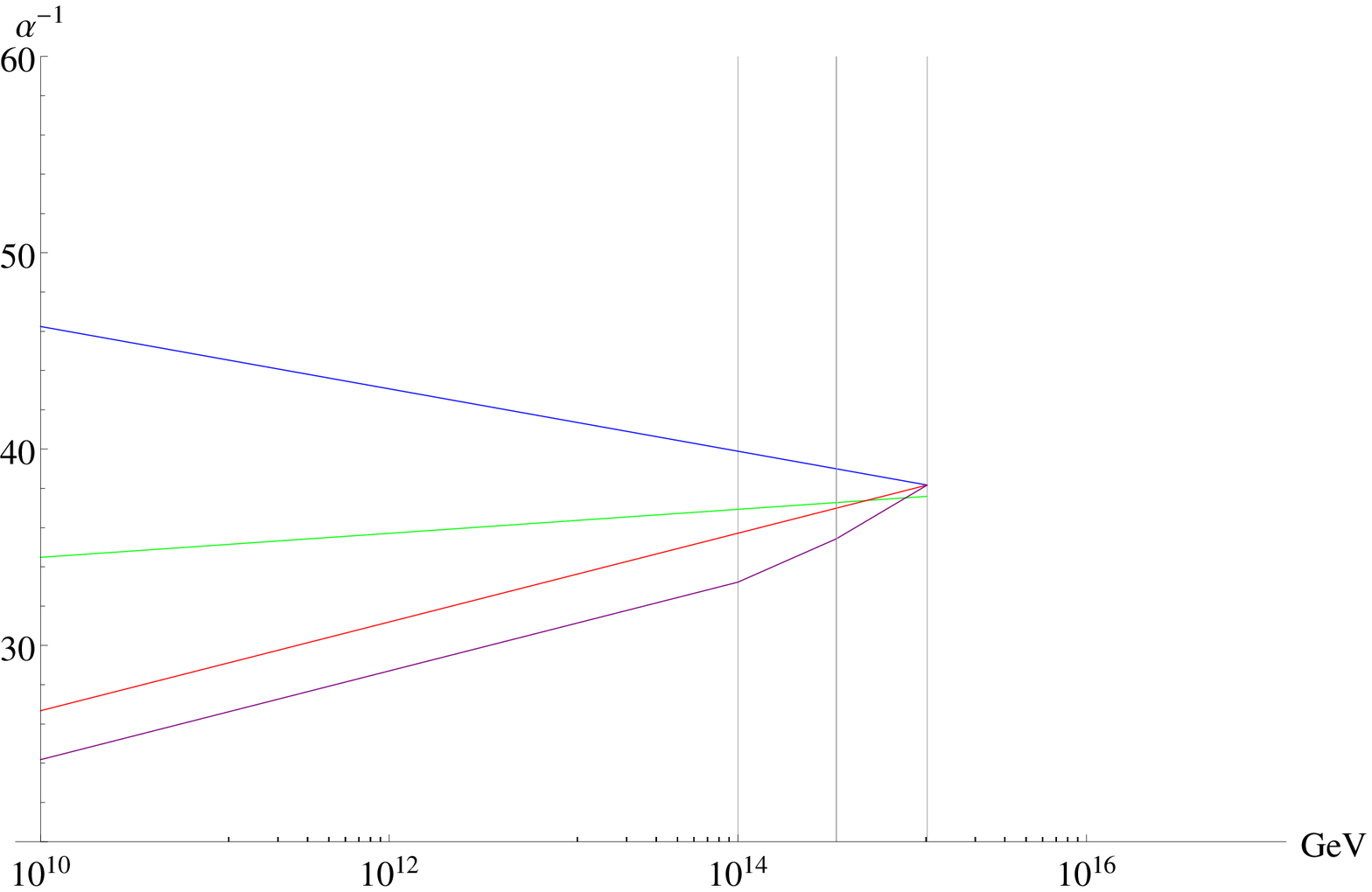}\\   
    \caption{Non-supersymmetric model with one and two intermediate scales. In the left plot we have $SU(3)$ in the visible sector and $SU(5)$ in the dark sector above the scale $M_X \approx 10^{15}$ GeV while the right plot shows $SU(3)$ in the visible sector and $SU(5)$ breaking to $SU(4)$ at a scale $M_J > M_I$ in the dark sector. Each graph displays the running coupling of the SM forces ($\alpha_1, \alpha_2, \alpha_3$) from top to bottom and that of the color force in the dark sector(the lowest line). 
    The value of the dark confinement scale is 3.2 GeV and 2.5 GeV respectively.}
    \label{fig:sub1}
  \end{minipage}\\[1em]
\end{figure}
                          One could examine a limitless number of such models in this context extending the number of breaking scales, however we can see that for almost any choice in the number of such scales and breaking chains in the VS, a model can be constructed which allows a dark confinement scale
                          through the effect of asymmetric symmetry breaking. In this sense it would be interesting to move on to developing a detailed model which resolves a significant number of other issues associated with GUTs in the VS and then adapt it to an ASB model in the pursuit of explaining dark matter also. 
                          A full theory of baryogenesis in the two sectors can also place strict limits on the size of these intermediate scales particularly in the case of baryogenesis via leptogenesis or GUT baryogenesis where the 
                          symmetry breaking scale can affect the amount of baryon number violation in the early universe.
                          In addition to these constraints we must also consider the current DM constraints on self interaction where the bullet cluster observation sets results on  self interaction for nucleon-nucleon like scattering in \cite{An:2009vq}. 
                          Such nucleon-like scattering has a cross section of $\sigma \sim 10^{-26}\;  {\rm cm}^2$  and can be compared to the upper bound of the DM self-interaction cross section $\le 10^{-23} \; {\rm cm}^2$ \cite{An:2009vq,Markevitch:2003at,Spergel:1999mh}. In the cases we have considered we were only concerned
                          with maintaining an SU(3) symmetry in the DS and so in many of these models the DM candidate only interacts with itself through short range strong forces and gravity.
                          Such neutral baryon dark matter particles are thus compatible with current detection limits.

                          \section{\bf Conclusions}\label{sec:Conclusions}
                         The similarity in the abundances of visible and dark matter leads us to suppose that their origin is not independent but rather that the production of both the number density of dark matter particles in the early universe and 
                         the dark matter mass is a result of processes that are deeply connected to the standard model. Mirror symmetric grand unified theories offer a plausible solution to both of these problems by supposing that
                          dark matter is a result of a hidden sector whose complexity is at least as large as our own yet is derived from an underlying theory that places the two components of the universe on exactly the same footing in the distant past. 
                         By assuming that the mass scale of dark matter comes from the confinement scale of a dark SU(3), this connection in masses
                         can be realized through grand unified theories. Where mirror symmetric models can give the similarity by having the dark sector be an exact copy of the SM, asymmetric symmetry breaking allows us to establish a much larger set
                         of theories of the universe which contain completely symmetric sectors whose GUT groups break to subgroups that are necessarily different allowing for the dynamics of visible and dark sectors to appear completely different in the low energy regime
                         while having the appealing concept of an underlying theory based on symmetries.
                         In this work we have expanded these asymmetric models to allow for SO(10) models to provide the similarity in visible and dark baryon mass by having multi-step breaking chains in mirror GUT scenarios give altered coupling constant evolution in the range between
                         intermediate scales and the GUT scale. We have demonstrated a specific $SO(10) \times SO(10)$ potential that breaks the symmetry asymmetrically to allow for such divergences in the coupling constant evolutions. The couplings, once gaining an altered value in the intermediate range, then run parallel all the way down to the low energy scale of the present day universe where the divergence of the QCD coupling in the two sectors occurs at 
                         separate but similar scales. This combined with the theory of ADM has the potential for a natural explanation of the apparently coincidental ratio $\Omega_{DM} \simeq 5 \Omega_{VM}$ that characterizes the matter of our universe.
                      
\acknowledgments{The author would like to thank Ray Volkas for helpful discussion and guidance. This work was supported in part by the Australian Research Council.}

                  \appendix
\section{\bf Scalar potentials for non-supersymmetric SO(10)$\times$SO(10) models}\label{sec:AppA}
In this appendix we expand on the potential discussed in Section \ref{sec:ASB}. We firstly consider the two representations of SO(10) independently. These are
the adjoint \textbf{45}, denoted by $\phi_{ij}$ which can be formed from the antisymmetric product of two fundamental representations, 
and the \textbf{54} which we label $\chi_{ij}$ which is formed from the completely symmetric product of two fundamentals.
The most general quartic potential for a rank two antisymmetric tensor in SO(10) is 
\begin{equation}
 -\frac{\mu^2}{2} {\phi}_{ij}{\phi}_{ji}  +\frac{\lambda}{4} ({\phi}_{ij}{\phi}_{ji})^2 + 
\frac{\alpha}{4}{\phi}_{ij}{\phi}_{jk}{\phi}_{kl}{\phi}_{li}.
\end{equation}
For this potential the symmetry breaking pattern is as follows. For $\lambda > 0$ and $\alpha > 0$ we have 
\begin{equation}
SO(10)  \rightarrow  SU(5) \times U(1),
\end{equation}
while for $\lambda > 0 $ and $\alpha < 0$ we find
\begin{equation}
SO(10)  \rightarrow  SO(8) \times U(1).
\end{equation}
In the case of the symmetric rank two representation we have a similar equation but with the added cubic term $Tr(\chi^3)$ so that the potential reads
\begin{equation}
 -\frac{\mu^2}{2} {\chi}_{ij}{\chi}_{ji}  +\frac{\lambda}{4} ({\chi}_{ij}{\chi}_{ji})^2 + 
\frac{\alpha}{4}{\chi}_{ij}{\chi}_{jk}{\chi}_{kl}{\chi}_{li} + \frac{\beta \mu }{3} {\chi}_{ij}{\chi}_{jk}{\chi}_{ki}.
\end{equation}
For this potential the parameter space is such that without the cubic term the possible breaking chains are, when $\lambda < 0$ and $\alpha < 0 $ the pattern is
 \begin{equation}
SO(10)  \rightarrow  SO(9),
\end{equation}
while for $\lambda > 0$ and $\alpha > 0 $ we have
 \begin{equation}
SO(10)  \rightarrow  SO(5) \times SO(5).
\end{equation}
For the parameter space where $\lambda > 0$ and $\alpha > 0 $ and the cubic term is non-zero we have
 \begin{equation}
SO(10)  \rightarrow  SO(10-n) \times SO(n),
\end{equation}
where for values of $\beta = 0$ we recover the above result in Eq. A6 and for $\beta > 0$, n increases as $\beta$ does until the breaking chain of Eq. A5 is recovered.
The generation of the potential in Eq. 18 then results from the addition of the two potentials given above in Eq. A1 and Eq. A4, 
as well as the analogue terms of the toy potential that mix the fields of the two sectors and the new non-trivial same-sector contractions 
afforded by the choice of the \textbf{45} and \textbf{54} representations.
Using this potential our numerical results align with the expected minima from the above potentials in the case where the cross terms and the additional cubic terms are sufficiently small. 
For the choice of parameter space where $\lambda_{\phi} > 0, \alpha_{\phi} > 0, \lambda_{\chi} > 0, \alpha_{\chi} > 0, \beta > 0 , \kappa_{\phi} > 0, \kappa_{\chi}> 0, c_2 > c_1 > 0,c_3 \ll c_2,c_4 \ll c_2,c_5 \ll c_2 $ we will obtain
a potential which breaks asymmetrically with the specific choice of breaking chain for each sector. This agrees with our numerical analysis where for a sample choice of parameters,
\begin{eqnarray}
& \lambda_{\phi} \simeq 1, \     
\kappa_{\phi} \simeq 0.75, \     
\kappa_{\chi}  \simeq 0.75, \
\lambda_{\chi} \simeq 1.6, & \nonumber \\    
& \mu_{\phi}  \simeq 1, \    
\mu_{\chi}  \simeq 1, \     
\alpha_{\phi}  \simeq 0.5, \      
\alpha_{\chi}  \simeq 1, & \nonumber \\    
& \beta_{\chi}  \simeq 0.35, \   
c_1  \simeq 0.25, \    
c_2  \simeq 0.75,  \ 
c_3  \simeq 0, \
c_4  \simeq 0,  \    
c_5  \simeq 0, & \nonumber 
\end{eqnarray}
we find that minimum preserves the VEVs 
\begin{eqnarray}
\braket{\phi_V} & = & 0.3
\begin{pmatrix}
0 & 1 & 0 & 0 & 0 & 0 & 0 & 0 & 0 & 0 \\
-1 & 0 & 0 & 0 & 0 & 0 & 0 & 0 & 0 & 0 \\
0 & 0 & 0 & 1 & 0 & 0 & 0 & 0 & 0 & 0 \\
0 & 0 & -1 & 0 & 0 & 0 & 0 & 0 & 0 & 0 \\
0 & 0 & 0 & 0 & 0 & 1 & 0 & 0 & 0 & 0 \\
0 & 0 & 0 & 0 & -1 & 0 & 0 & 0 & 0 & 0 \\
0 & 0 & 0 & 0 & 0 & 0 & 0 & 1 & 0 & 0 \\
0 & 0 & 0 & 0 & 0 & 0 & -1 & 0 & 0 & 0 \\
0 & 0 & 0 & 0 & 0 & 0 & 0 & 0 & 0 & 1 \\
0 & 0 & 0 & 0 & 0 & 0 & 0 & 0 & -1 & 0 \\
 \end{pmatrix} \nonumber \\
\braket{\phi_D} & = & 0 \nonumber \\
\braket{\chi_V} & = & 0 \nonumber \\
\braket{\chi_D} & = & 0.3
 \begin{pmatrix}
1 & 0 & 0 & 0 & 0 & 0 & 0 & 0 & 0 & 0 \\
0 & 1 & 0 & 0 & 0 & 0 & 0 & 0 & 0 & 0 \\
0 & 0 & 1 & 0 & 0 & 0 & 0 & 0 & 0 & 0 \\
0 & 0 & 0 & 1 & 0 & 0 & 0 & 0 & 0 & 0 \\
0 & 0 & 0 & 0 & 1 & 0 & 0 & 0 & 0 & 0 \\
0 & 0 & 0 & 0 & 0 & 1 & 0 & 0 & 0 & 0 \\
0 & 0 & 0 & 0 & 0 & 0 & -6/4 & 0 & 0 & 0 \\
0 & 0 & 0 & 0 & 0 & 0 & 0 & -6/4 & 0 & 0 \\
0 & 0 & 0 & 0 & 0 & 0 & 0 & 0 & -6/4 & 0 \\
0 & 0 & 0 & 0 & 0 & 0 & 0 & 0 & 0 & -6/4 \\
 \end{pmatrix},
\end{eqnarray}
which breaks the symmetry according to
 \begin{equation}
SO(10)_V \times SO(10)_D  \rightarrow  [SU(4)\times SU(2) \times SU(2)]_V \times [SU(5) \times U(1)]_D.
\end{equation}
The analysis discussed here describes just the first step in asymmetrically breaking an SO(10) mirror symmetric potential to different subgroups for each sector and at different energy scales. 
While many other possible breaking chains that have been discussed in this paper could be analyzed, we leave such work to future efforts to create a detailed model of an SO(10) GUT model where the choice of representations aligns with choices 
for fermion mass generation models and considerations of minimality. Due to the complexity in analyzing such Higgs potentials for large gauge groups we content 
ourselves at the present juncture with the demonstration of the versatility of such asymmetric symmetry breaking in the context of GUT models. 
With this specific example and the principles given in the toy model, many of the other breaking chains could be realized in potentials constructed in a like manner.


\bibliography{dmfromI.bib}

\begin{thebibliography}{45}%
\makeatletter
\providecommand \@ifxundefined [1]{%
 \@ifx{#1\undefined}
}%
\providecommand \@ifnum [1]{%
 \ifnum #1\expandafter \@firstoftwo
 \else \expandafter \@secondoftwo
 \fi
}%
\providecommand \@ifx [1]{%
 \ifx #1\expandafter \@firstoftwo
 \else \expandafter \@secondoftwo
 \fi
}%
\providecommand \natexlab [1]{#1}%
\providecommand \enquote  [1]{``#1''}%
\providecommand \bibnamefont  [1]{#1}%
\providecommand \bibfnamefont [1]{#1}%
\providecommand \citenamefont [1]{#1}%
\providecommand \href@noop [0]{\@secondoftwo}%
\providecommand \href [0]{\begingroup \@sanitize@url \@href}%
\providecommand \@href[1]{\@@startlink{#1}\@@href}%
\providecommand \@@href[1]{\endgroup#1\@@endlink}%
\providecommand \@sanitize@url [0]{\catcode `\\12\catcode `\$12\catcode
  `\&12\catcode `\#12\catcode `\^12\catcode `\_12\catcode `\%12\relax}%
\providecommand \@@startlink[1]{}%
\providecommand \@@endlink[0]{}%
\providecommand \url  [0]{\begingroup\@sanitize@url \@url }%
\providecommand \@url [1]{\endgroup\@href {#1}{\urlprefix }}%
\providecommand \urlprefix  [0]{URL }%
\providecommand \Eprint [0]{\href }%
\providecommand \doibase [0]{http://dx.doi.org/}%
\providecommand \selectlanguage [0]{\@gobble}%
\providecommand \bibinfo  [0]{\@secondoftwo}%
\providecommand \bibfield  [0]{\@secondoftwo}%
\providecommand \translation [1]{[#1]}%
\providecommand \BibitemOpen [0]{}%
\providecommand \bibitemStop [0]{}%
\providecommand \bibitemNoStop [0]{.\EOS\space}%
\providecommand \EOS [0]{\spacefactor3000\relax}%
\providecommand \BibitemShut  [1]{\csname bibitem#1\endcsname}%
\let\auto@bib@innerbib\@empty
\bibitem [{\citenamefont {Davoudiasl}\ and\ \citenamefont
  {Mohapatra}(2012)}]{Davoudiasl:2012uw}%
  \BibitemOpen
  \bibfield  {author} {\bibinfo {author} {\bibfnamefont {H.}~\bibnamefont
  {Davoudiasl}}\ and\ \bibinfo {author} {\bibfnamefont {R.~N.}\ \bibnamefont
  {Mohapatra}},\ }\href {\doibase 10.1088/1367-2630/14/9/095011} {\bibfield
  {journal} {\bibinfo  {journal} {New J.Phys.}\ }\textbf {\bibinfo {volume}
  {14}},\ \bibinfo {pages} {095011} (\bibinfo {year} {2012})},\ \Eprint
  {http://arxiv.org/abs/1203.1247} {arXiv:1203.1247 [hep-ph]} \BibitemShut
  {NoStop}%
\bibitem [{\citenamefont {Petraki}\ and\ \citenamefont
  {Volkas}(2013)}]{Petraki:2013wwa}%
  \BibitemOpen
  \bibfield  {author} {\bibinfo {author} {\bibfnamefont {K.}~\bibnamefont
  {Petraki}}\ and\ \bibinfo {author} {\bibfnamefont {R.~R.}\ \bibnamefont
  {Volkas}},\ }\href {\doibase 10.1142/S0217751X13300287} {\bibfield  {journal}
  {\bibinfo  {journal} {Int.J.Mod.Phys.}\ }\textbf {\bibinfo {volume} {A28}},\
  \bibinfo {pages} {1330028} (\bibinfo {year} {2013})},\ \Eprint
  {http://arxiv.org/abs/1305.4939} {arXiv:1305.4939 [hep-ph]} \BibitemShut
  {NoStop}%
\bibitem [{\citenamefont {Zurek}(2014)}]{Zurek:2013wia}%
  \BibitemOpen
  \bibfield  {author} {\bibinfo {author} {\bibfnamefont {K.~M.}\ \bibnamefont
  {Zurek}},\ }\href {\doibase 10.1016/j.physrep.2013.12.001} {\bibfield
  {journal} {\bibinfo  {journal} {Phys.Rept.}\ }\textbf {\bibinfo {volume}
  {537}},\ \bibinfo {pages} {91} (\bibinfo {year} {2014})},\ \Eprint
  {http://arxiv.org/abs/1308.0338} {arXiv:1308.0338 [hep-ph]} \BibitemShut
  {NoStop}%
\bibitem [{\citenamefont {Lonsdale}\ and\ \citenamefont
  {Volkas}(2014)}]{Lonsdale:2014wwa}%
  \BibitemOpen
  \bibfield  {author} {\bibinfo {author} {\bibfnamefont {S.~J.}\ \bibnamefont
  {Lonsdale}}\ and\ \bibinfo {author} {\bibfnamefont {R.~R.}\ \bibnamefont
  {Volkas}},\ }\href {\doibase 10.1103/PhysRevD.90.083501} {\bibfield
  {journal} {\bibinfo  {journal} {Phys.Rev.}\ }\textbf {\bibinfo {volume}
  {D90}},\ \bibinfo {pages} {083501} (\bibinfo {year} {2014})},\ \Eprint
  {http://arxiv.org/abs/1407.4192} {arXiv:1407.4192 [hep-ph]} \BibitemShut
  {NoStop}%
\bibitem [{\citenamefont {Lee}\ and\ \citenamefont {Yang}(1956)}]{Lee:1956qn}%
  \BibitemOpen
  \bibfield  {author} {\bibinfo {author} {\bibfnamefont {T.}~\bibnamefont
  {Lee}}\ and\ \bibinfo {author} {\bibfnamefont {C.-N.}\ \bibnamefont {Yang}},\
  }\href {\doibase 10.1103/PhysRev.104.254} {\bibfield  {journal} {\bibinfo
  {journal} {Phys.Rev.}\ }\textbf {\bibinfo {volume} {104}},\ \bibinfo {pages}
  {254} (\bibinfo {year} {1956})}\BibitemShut {NoStop}%
\bibitem [{\citenamefont {Kobzarev}\ \emph {et~al.}(1966)\citenamefont
  {Kobzarev}, \citenamefont {Okun},\ and\ \citenamefont
  {Pomeranchuk}}]{Kobzarev:1966}%
  \BibitemOpen
  \bibfield  {author} {\bibinfo {author} {\bibfnamefont {I.}~\bibnamefont
  {Kobzarev}}, \bibinfo {author} {\bibfnamefont {L.}~\bibnamefont {Okun}}, \
  and\ \bibinfo {author} {\bibfnamefont {I.}~\bibnamefont {Pomeranchuk}},\
  }\href@noop {} {\bibfield  {journal} {\bibinfo  {journal} {Sov.J.Nucl.Phys.}\
  }\textbf {\bibinfo {volume} {3}},\ \bibinfo {pages} {837} (\bibinfo {year}
  {1966})}\BibitemShut {NoStop}%
\bibitem [{\citenamefont {Pavsic}(1974)}]{Pavsic:1974rq}%
  \BibitemOpen
  \bibfield  {author} {\bibinfo {author} {\bibfnamefont {M.}~\bibnamefont
  {Pavsic}},\ }\href {\doibase 10.1007/BF01810695} {\bibfield  {journal}
  {\bibinfo  {journal} {Int.J.Theor.Phys.}\ }\textbf {\bibinfo {volume} {9}},\
  \bibinfo {pages} {229} (\bibinfo {year} {1974})},\ \Eprint
  {http://arxiv.org/abs/hep-ph/0105344} {arXiv:hep-ph/0105344 [hep-ph]}
  \BibitemShut {NoStop}%
\bibitem [{\citenamefont {Blinnikov}\ and\ \citenamefont
  {Khlopov}(1982)}]{Blinnikov:1982eh}%
  \BibitemOpen
  \bibfield  {author} {\bibinfo {author} {\bibfnamefont {S.}~\bibnamefont
  {Blinnikov}}\ and\ \bibinfo {author} {\bibfnamefont {M.~Y.}\ \bibnamefont
  {Khlopov}},\ }\href@noop {} {\bibfield  {journal} {\bibinfo  {journal}
  {Sov.J.Nucl.Phys.}\ }\textbf {\bibinfo {volume} {36}},\ \bibinfo {pages}
  {472} (\bibinfo {year} {1982})}\BibitemShut {NoStop}%
\bibitem [{\citenamefont {S.Blinnikov}\ and\ \citenamefont
  {M.Khlopov}(1983)}]{Blinnikov:1983}%
  \BibitemOpen
  \bibfield  {author} {\bibinfo {author} {\bibnamefont {S.Blinnikov}}\ and\
  \bibinfo {author} {\bibnamefont {M.Khlopov}},\ }\href@noop {} {\bibfield
  {journal} {\bibinfo  {journal} {Sov.Astron.Lett.}\ }\textbf {\bibinfo
  {volume} {27}},\ \bibinfo {pages} {371} (\bibinfo {year} {1983})}\BibitemShut
  {NoStop}%
\bibitem [{\citenamefont {Foot}\ \emph {et~al.}(1991)\citenamefont {Foot},
  \citenamefont {Lew},\ and\ \citenamefont {Volkas}}]{Foot:1991bp}%
  \BibitemOpen
  \bibfield  {author} {\bibinfo {author} {\bibfnamefont {R.}~\bibnamefont
  {Foot}}, \bibinfo {author} {\bibfnamefont {H.}~\bibnamefont {Lew}}, \ and\
  \bibinfo {author} {\bibfnamefont {R.}~\bibnamefont {Volkas}},\ }\href
  {\doibase 10.1016/0370-2693(91)91013-L} {\bibfield  {journal} {\bibinfo
  {journal} {Phys.Lett.}\ }\textbf {\bibinfo {volume} {B272}},\ \bibinfo
  {pages} {67} (\bibinfo {year} {1991})}\BibitemShut {NoStop}%
\bibitem [{\citenamefont {Foot}\ \emph {et~al.}(1992)\citenamefont {Foot},
  \citenamefont {Lew},\ and\ \citenamefont {Volkas}}]{Foot:1991py}%
  \BibitemOpen
  \bibfield  {author} {\bibinfo {author} {\bibfnamefont {R.}~\bibnamefont
  {Foot}}, \bibinfo {author} {\bibfnamefont {H.}~\bibnamefont {Lew}}, \ and\
  \bibinfo {author} {\bibfnamefont {R.}~\bibnamefont {Volkas}},\ }\href
  {\doibase 10.1142/S0217732392004031} {\bibfield  {journal} {\bibinfo
  {journal} {Mod.Phys.Lett.}\ }\textbf {\bibinfo {volume} {A7}},\ \bibinfo
  {pages} {2567} (\bibinfo {year} {1992})}\BibitemShut {NoStop}%
\bibitem [{\citenamefont {Foot}\ and\ \citenamefont
  {Volkas}(1995)}]{Foot:1995pa}%
  \BibitemOpen
  \bibfield  {author} {\bibinfo {author} {\bibfnamefont {R.}~\bibnamefont
  {Foot}}\ and\ \bibinfo {author} {\bibfnamefont {R.~R.}\ \bibnamefont
  {Volkas}},\ }\href {\doibase 10.1103/PhysRevD.52.6595} {\bibfield  {journal}
  {\bibinfo  {journal} {Phys.Rev.}\ }\textbf {\bibinfo {volume} {D52}},\
  \bibinfo {pages} {6595} (\bibinfo {year} {1995})},\ \Eprint
  {http://arxiv.org/abs/hep-ph/9505359} {arXiv:hep-ph/9505359 [hep-ph]}
  \BibitemShut {NoStop}%
\bibitem [{\citenamefont {Berezhiani}\ and\ \citenamefont
  {Mohapatra}(1995)}]{Berezhiani:1995yi}%
  \BibitemOpen
  \bibfield  {author} {\bibinfo {author} {\bibfnamefont {Z.~G.}\ \bibnamefont
  {Berezhiani}}\ and\ \bibinfo {author} {\bibfnamefont {R.~N.}\ \bibnamefont
  {Mohapatra}},\ }\href {\doibase 10.1103/PhysRevD.52.6607} {\bibfield
  {journal} {\bibinfo  {journal} {Phys.Rev.}\ }\textbf {\bibinfo {volume}
  {D52}},\ \bibinfo {pages} {6607} (\bibinfo {year} {1995})},\ \Eprint
  {http://arxiv.org/abs/hep-ph/9505385} {arXiv:hep-ph/9505385 [hep-ph]}
  \BibitemShut {NoStop}%
\bibitem [{\citenamefont {Foot}\ \emph {et~al.}(2000)\citenamefont {Foot},
  \citenamefont {Lew},\ and\ \citenamefont {Volkas}}]{Foot:2000tp}%
  \BibitemOpen
  \bibfield  {author} {\bibinfo {author} {\bibfnamefont {R.}~\bibnamefont
  {Foot}}, \bibinfo {author} {\bibfnamefont {H.}~\bibnamefont {Lew}}, \ and\
  \bibinfo {author} {\bibfnamefont {R.}~\bibnamefont {Volkas}},\ }\href
  {\doibase 10.1088/1126-6708/2000/07/032} {\bibfield  {journal} {\bibinfo
  {journal} {JHEP}\ }\textbf {\bibinfo {volume} {0007}},\ \bibinfo {pages}
  {032} (\bibinfo {year} {2000})},\ \Eprint
  {http://arxiv.org/abs/hep-ph/0006027} {arXiv:hep-ph/0006027 [hep-ph]}
  \BibitemShut {NoStop}%
\bibitem [{\citenamefont {Berezhiani}\ \emph {et~al.}(2001)\citenamefont
  {Berezhiani}, \citenamefont {Comelli},\ and\ \citenamefont
  {Villante}}]{Berezhiani:2000gw}%
  \BibitemOpen
  \bibfield  {author} {\bibinfo {author} {\bibfnamefont {Z.}~\bibnamefont
  {Berezhiani}}, \bibinfo {author} {\bibfnamefont {D.}~\bibnamefont {Comelli}},
  \ and\ \bibinfo {author} {\bibfnamefont {F.~L.}\ \bibnamefont {Villante}},\
  }\href {\doibase 10.1016/S0370-2693(01)00217-9} {\bibfield  {journal}
  {\bibinfo  {journal} {Phys.Lett.}\ }\textbf {\bibinfo {volume} {B503}},\
  \bibinfo {pages} {362} (\bibinfo {year} {2001})},\ \Eprint
  {http://arxiv.org/abs/hep-ph/0008105} {arXiv:hep-ph/0008105 [hep-ph]}
  \BibitemShut {NoStop}%
\bibitem [{\citenamefont {Ignatiev}\ and\ \citenamefont
  {Volkas}(2003)}]{Ignatiev:2003js}%
  \BibitemOpen
  \bibfield  {author} {\bibinfo {author} {\bibfnamefont {A.~Y.}\ \bibnamefont
  {Ignatiev}}\ and\ \bibinfo {author} {\bibfnamefont {R.~R.}\ \bibnamefont
  {Volkas}},\ }\href {\doibase 10.1103/PhysRevD.68.023518} {\bibfield
  {journal} {\bibinfo  {journal} {Phys.Rev.}\ }\textbf {\bibinfo {volume}
  {D68}},\ \bibinfo {pages} {023518} (\bibinfo {year} {2003})},\ \Eprint
  {http://arxiv.org/abs/hep-ph/0304260} {arXiv:hep-ph/0304260 [hep-ph]}
  \BibitemShut {NoStop}%
\bibitem [{\citenamefont {Foot}\ and\ \citenamefont
  {Volkas}(2003)}]{Foot:2003jt}%
  \BibitemOpen
  \bibfield  {author} {\bibinfo {author} {\bibfnamefont {R.}~\bibnamefont
  {Foot}}\ and\ \bibinfo {author} {\bibfnamefont {R.~R.}\ \bibnamefont
  {Volkas}},\ }\href {\doibase 10.1103/PhysRevD.68.021304} {\bibfield
  {journal} {\bibinfo  {journal} {Phys.Rev.}\ }\textbf {\bibinfo {volume}
  {D68}},\ \bibinfo {pages} {021304} (\bibinfo {year} {2003})},\ \Eprint
  {http://arxiv.org/abs/hep-ph/0304261} {arXiv:hep-ph/0304261 [hep-ph]}
  \BibitemShut {NoStop}%
\bibitem [{\citenamefont {Foot}\ and\ \citenamefont
  {Volkas}(2004)}]{Foot:2004pq}%
  \BibitemOpen
  \bibfield  {author} {\bibinfo {author} {\bibfnamefont {R.}~\bibnamefont
  {Foot}}\ and\ \bibinfo {author} {\bibfnamefont {R.~R.}\ \bibnamefont
  {Volkas}},\ }\href {\doibase 10.1103/PhysRevD.69.123510} {\bibfield
  {journal} {\bibinfo  {journal} {Phys.Rev.}\ }\textbf {\bibinfo {volume}
  {D69}},\ \bibinfo {pages} {123510} (\bibinfo {year} {2004})},\ \Eprint
  {http://arxiv.org/abs/hep-ph/0402267} {arXiv:hep-ph/0402267 [hep-ph]}
  \BibitemShut {NoStop}%
\bibitem [{\citenamefont {Berezhiani}\ \emph {et~al.}(2005)\citenamefont
  {Berezhiani}, \citenamefont {Ciarcelluti}, \citenamefont {Comelli},\ and\
  \citenamefont {Villante}}]{Berezhiani:2003wj}%
  \BibitemOpen
  \bibfield  {author} {\bibinfo {author} {\bibfnamefont {Z.}~\bibnamefont
  {Berezhiani}}, \bibinfo {author} {\bibfnamefont {P.}~\bibnamefont
  {Ciarcelluti}}, \bibinfo {author} {\bibfnamefont {D.}~\bibnamefont
  {Comelli}}, \ and\ \bibinfo {author} {\bibfnamefont {F.~L.}\ \bibnamefont
  {Villante}},\ }\href {\doibase 10.1142/S0218271805005165} {\bibfield
  {journal} {\bibinfo  {journal} {Int.J.Mod.Phys.}\ }\textbf {\bibinfo {volume}
  {D14}},\ \bibinfo {pages} {107} (\bibinfo {year} {2005})},\ \Eprint
  {http://arxiv.org/abs/astro-ph/0312605} {arXiv:astro-ph/0312605 [astro-ph]}
  \BibitemShut {NoStop}%
\bibitem [{\citenamefont
  {Ciarcelluti}(2005{\natexlab{a}})}]{Ciarcelluti:2004ik}%
  \BibitemOpen
  \bibfield  {author} {\bibinfo {author} {\bibfnamefont {P.}~\bibnamefont
  {Ciarcelluti}},\ }\href {\doibase 10.1142/S0218271805006213} {\bibfield
  {journal} {\bibinfo  {journal} {Int.J.Mod.Phys.}\ }\textbf {\bibinfo {volume}
  {D14}},\ \bibinfo {pages} {187} (\bibinfo {year} {2005}{\natexlab{a}})},\
  \Eprint {http://arxiv.org/abs/astro-ph/0409630} {arXiv:astro-ph/0409630
  [astro-ph]} \BibitemShut {NoStop}%
\bibitem [{\citenamefont
  {Ciarcelluti}(2005{\natexlab{b}})}]{Ciarcelluti:2004ip}%
  \BibitemOpen
  \bibfield  {author} {\bibinfo {author} {\bibfnamefont {P.}~\bibnamefont
  {Ciarcelluti}},\ }\href {\doibase 10.1142/S0218271805006225} {\bibfield
  {journal} {\bibinfo  {journal} {Int.J.Mod.Phys.}\ }\textbf {\bibinfo {volume}
  {D14}},\ \bibinfo {pages} {223} (\bibinfo {year} {2005}{\natexlab{b}})},\
  \Eprint {http://arxiv.org/abs/astro-ph/0409633} {arXiv:astro-ph/0409633
  [astro-ph]} \BibitemShut {NoStop}%
\bibitem [{\citenamefont {Foot}(2014)}]{Foot:2014mia}%
  \BibitemOpen
  \bibfield  {author} {\bibinfo {author} {\bibfnamefont {R.}~\bibnamefont
  {Foot}},\ }\href {\doibase 10.1142/S0217751X14300130} {\bibfield  {journal}
  {\bibinfo  {journal} {Int.J.Mod.Phys.}\ }\textbf {\bibinfo {volume} {A29}},\
  \bibinfo {pages} {1430013} (\bibinfo {year} {2014})},\ \Eprint
  {http://arxiv.org/abs/1401.3965} {arXiv:1401.3965 [astro-ph.CO]} \BibitemShut
  {NoStop}%
\bibitem [{\citenamefont {Gu}(2014)}]{Gu:2014nga}%
  \BibitemOpen
  \bibfield  {author} {\bibinfo {author} {\bibfnamefont {P.-H.}\ \bibnamefont
  {Gu}},\ }\href@noop {} {\  (\bibinfo {year} {2014})},\ \Eprint
  {http://arxiv.org/abs/1410.5759} {arXiv:1410.5759 [hep-ph]} \BibitemShut
  {NoStop}%
\bibitem [{\citenamefont {Bai}\ and\ \citenamefont
  {Schwaller}(2014)}]{Bai:2013xga}%
  \BibitemOpen
  \bibfield  {author} {\bibinfo {author} {\bibfnamefont {Y.}~\bibnamefont
  {Bai}}\ and\ \bibinfo {author} {\bibfnamefont {P.}~\bibnamefont
  {Schwaller}},\ }\href {\doibase 10.1103/PhysRevD.89.063522} {\bibfield
  {journal} {\bibinfo  {journal} {Phys.Rev.}\ }\textbf {\bibinfo {volume}
  {D89}},\ \bibinfo {pages} {063522} (\bibinfo {year} {2014})},\ \Eprint
  {http://arxiv.org/abs/1306.4676} {arXiv:1306.4676 [hep-ph]} \BibitemShut
  {NoStop}%
\bibitem [{\citenamefont {Barr}\ and\ \citenamefont
  {Chen}(2013)}]{Barr:2013tea}%
  \BibitemOpen
  \bibfield  {author} {\bibinfo {author} {\bibfnamefont {S.}~\bibnamefont
  {Barr}}\ and\ \bibinfo {author} {\bibfnamefont {H.-Y.}\ \bibnamefont
  {Chen}},\ }\href {\doibase 10.1007/JHEP10(2013)129} {\bibfield  {journal}
  {\bibinfo  {journal} {JHEP}\ }\textbf {\bibinfo {volume} {1310}},\ \bibinfo
  {pages} {129} (\bibinfo {year} {2013})},\ \Eprint
  {http://arxiv.org/abs/1309.0020} {arXiv:1309.0020 [hep-ph]} \BibitemShut
  {NoStop}%
\bibitem [{\citenamefont {Ma}(2013)}]{Ma:2013nga}%
  \BibitemOpen
  \bibfield  {author} {\bibinfo {author} {\bibfnamefont {E.}~\bibnamefont
  {Ma}},\ }\href@noop {} {\  (\bibinfo {year} {2013})},\ \Eprint
  {http://arxiv.org/abs/1307.7064} {arXiv:1307.7064} \BibitemShut {NoStop}%
\bibitem [{\citenamefont {Newstead}\ and\ \citenamefont
  {TerBeek}(2014)}]{Newstead:2014jva}%
  \BibitemOpen
  \bibfield  {author} {\bibinfo {author} {\bibfnamefont {J.~L.}\ \bibnamefont
  {Newstead}}\ and\ \bibinfo {author} {\bibfnamefont {R.~H.}\ \bibnamefont
  {TerBeek}},\ }\href@noop {} {\  (\bibinfo {year} {2014})},\ \Eprint
  {http://arxiv.org/abs/1405.7427} {arXiv:1405.7427 [hep-ph]} \BibitemShut
  {NoStop}%
\bibitem [{\citenamefont {Boddy}\ \emph {et~al.}(2014)\citenamefont {Boddy},
  \citenamefont {Feng}, \citenamefont {Kaplinghat},\ and\ \citenamefont
  {Tait}}]{Boddy:2014yra}%
  \BibitemOpen
  \bibfield  {author} {\bibinfo {author} {\bibfnamefont {K.~K.}\ \bibnamefont
  {Boddy}}, \bibinfo {author} {\bibfnamefont {J.~L.}\ \bibnamefont {Feng}},
  \bibinfo {author} {\bibfnamefont {M.}~\bibnamefont {Kaplinghat}}, \ and\
  \bibinfo {author} {\bibfnamefont {T.~M.~P.}\ \bibnamefont {Tait}},\
  }\href@noop {} {\  (\bibinfo {year} {2014})},\ \Eprint
  {http://arxiv.org/abs/1402.3629} {arXiv:1402.3629 [hep-ph]} \BibitemShut
  {NoStop}%
\bibitem [{\citenamefont {Yamanaka}\ \emph {et~al.}(2014)\citenamefont
  {Yamanaka}, \citenamefont {Fujibayashi}, \citenamefont {Gongyo},\ and\
  \citenamefont {Iida}}]{Yamanaka:2014pva}%
  \BibitemOpen
  \bibfield  {author} {\bibinfo {author} {\bibfnamefont {N.}~\bibnamefont
  {Yamanaka}}, \bibinfo {author} {\bibfnamefont {S.}~\bibnamefont
  {Fujibayashi}}, \bibinfo {author} {\bibfnamefont {S.}~\bibnamefont {Gongyo}},
  \ and\ \bibinfo {author} {\bibfnamefont {H.}~\bibnamefont {Iida}},\
  }\href@noop {} {\  (\bibinfo {year} {2014})},\ \Eprint
  {http://arxiv.org/abs/1411.2172} {arXiv:1411.2172 [hep-ph]} \BibitemShut
  {NoStop}%
\bibitem [{\citenamefont {Higaki}\ \emph {et~al.}(2013)\citenamefont {Higaki},
  \citenamefont {Jeong},\ and\ \citenamefont {Takahashi}}]{Higaki:2013vuv}%
  \BibitemOpen
  \bibfield  {author} {\bibinfo {author} {\bibfnamefont {T.}~\bibnamefont
  {Higaki}}, \bibinfo {author} {\bibfnamefont {K.~S.}\ \bibnamefont {Jeong}}, \
  and\ \bibinfo {author} {\bibfnamefont {F.}~\bibnamefont {Takahashi}},\ }\href
  {\doibase 10.1088/1475-7516/2013/08/031} {\bibfield  {journal} {\bibinfo
  {journal} {JCAP}\ }\textbf {\bibinfo {volume} {1308}},\ \bibinfo {pages}
  {031} (\bibinfo {year} {2013})},\ \Eprint {http://arxiv.org/abs/1302.2516}
  {arXiv:1302.2516 [hep-ph]} \BibitemShut {NoStop}%
\bibitem [{\citenamefont {Tavartkiladze}(2014)}]{Tavartkiladze:2014lla}%
  \BibitemOpen
  \bibfield  {author} {\bibinfo {author} {\bibfnamefont {Z.}~\bibnamefont
  {Tavartkiladze}},\ }\href@noop {} {\  (\bibinfo {year} {2014})},\ \Eprint
  {http://arxiv.org/abs/1403.0025} {arXiv:1403.0025 [hep-ph]} \BibitemShut
  {NoStop}%
\bibitem [{\citenamefont {Ross}(1985)}]{Ross:1985ai}%
  \BibitemOpen
  \bibfield  {author} {\bibinfo {author} {\bibfnamefont {G.~G.}\ \bibnamefont
  {Ross}},\ }\href@noop {} {\emph {\bibinfo {title} {Grand Unified Theories}}}\
  (\bibinfo {year} {1985})\ p.\ \bibinfo {pages} {229}\BibitemShut {NoStop}%
\bibitem [{\citenamefont {Michel}(1980)}]{Michel:1980pc}%
  \BibitemOpen
  \bibfield  {author} {\bibinfo {author} {\bibfnamefont {L.}~\bibnamefont
  {Michel}},\ }\href {\doibase 10.1103/RevModPhys.52.617} {\bibfield  {journal}
  {\bibinfo  {journal} {Rev.Mod.Phys.}\ }\textbf {\bibinfo {volume} {52}},\
  \bibinfo {pages} {617} (\bibinfo {year} {1980})}\BibitemShut {NoStop}%
\bibitem [{\citenamefont {Slansky}(1981)}]{Slansky:1981yr}%
  \BibitemOpen
  \bibfield  {author} {\bibinfo {author} {\bibfnamefont {R.}~\bibnamefont
  {Slansky}},\ }\href {\doibase 10.1016/0370-1573(81)90092-2} {\bibfield
  {journal} {\bibinfo  {journal} {Phys.Rept.}\ }\textbf {\bibinfo {volume}
  {79}},\ \bibinfo {pages} {1} (\bibinfo {year} {1981})}\BibitemShut {NoStop}%
\bibitem [{\citenamefont {Pati}\ and\ \citenamefont
  {Salam}(1973)}]{Pati:1973rp}%
  \BibitemOpen
  \bibfield  {author} {\bibinfo {author} {\bibfnamefont {J.~C.}\ \bibnamefont
  {Pati}}\ and\ \bibinfo {author} {\bibfnamefont {A.}~\bibnamefont {Salam}},\
  }\href {\doibase 10.1103/PhysRevLett.31.661} {\bibfield  {journal} {\bibinfo
  {journal} {Phys.Rev.Lett.}\ }\textbf {\bibinfo {volume} {31}},\ \bibinfo
  {pages} {661} (\bibinfo {year} {1973})}\BibitemShut {NoStop}%
\bibitem [{\citenamefont {Georgi}\ and\ \citenamefont
  {Glashow}(1974)}]{Georgi:1974sy}%
  \BibitemOpen
  \bibfield  {author} {\bibinfo {author} {\bibfnamefont {H.}~\bibnamefont
  {Georgi}}\ and\ \bibinfo {author} {\bibfnamefont {S.}~\bibnamefont
  {Glashow}},\ }\href {\doibase 10.1103/PhysRevLett.32.438} {\bibfield
  {journal} {\bibinfo  {journal} {Phys.Rev.Lett.}\ }\textbf {\bibinfo {volume}
  {32}},\ \bibinfo {pages} {438} (\bibinfo {year} {1974})}\BibitemShut
  {NoStop}%
\bibitem [{\citenamefont {Georgi}(1982)}]{Georgi:1982jb}%
  \BibitemOpen
  \bibfield  {author} {\bibinfo {author} {\bibfnamefont {H.}~\bibnamefont
  {Georgi}},\ }\href@noop {} {\bibfield  {journal} {\bibinfo  {journal}
  {Front.Phys.}\ }\textbf {\bibinfo {volume} {54}},\ \bibinfo {pages} {1}
  (\bibinfo {year} {1982})}\BibitemShut {NoStop}%
\bibitem [{\citenamefont {Wu}(1982)}]{Wu:1981eb}%
  \BibitemOpen
  \bibfield  {author} {\bibinfo {author} {\bibfnamefont {D.-d.}\ \bibnamefont
  {Wu}},\ }\href {\doibase 10.1016/0550-3213(82)90358-3} {\bibfield  {journal}
  {\bibinfo  {journal} {Nucl.Phys.}\ }\textbf {\bibinfo {volume} {B199}},\
  \bibinfo {pages} {523} (\bibinfo {year} {1982})}\BibitemShut {NoStop}%
\bibitem [{\citenamefont {Dorsner}\ \emph {et~al.}(2006)\citenamefont
  {Dorsner}, \citenamefont {Fileviez~Perez},\ and\ \citenamefont
  {Gonzalez~Felipe}}]{Dorsner:2005ii}%
  \BibitemOpen
  \bibfield  {author} {\bibinfo {author} {\bibfnamefont {I.}~\bibnamefont
  {Dorsner}}, \bibinfo {author} {\bibfnamefont {P.}~\bibnamefont
  {Fileviez~Perez}}, \ and\ \bibinfo {author} {\bibfnamefont {R.}~\bibnamefont
  {Gonzalez~Felipe}},\ }\href {\doibase 10.1016/j.nuclphysb.2006.05.006}
  {\bibfield  {journal} {\bibinfo  {journal} {Nucl.Phys.}\ }\textbf {\bibinfo
  {volume} {B747}},\ \bibinfo {pages} {312} (\bibinfo {year} {2006})},\ \Eprint
  {http://arxiv.org/abs/hep-ph/0512068} {arXiv:hep-ph/0512068 [hep-ph]}
  \BibitemShut {NoStop}%
\bibitem [{\citenamefont {Senjanovic}(2010)}]{Senjanovic:2009}%
  \BibitemOpen
  \bibfield  {author} {\bibinfo {author} {\bibfnamefont {G.}~\bibnamefont
  {Senjanovic}},\ }\href {\doibase 10.1063/1.3327552} {\bibfield  {journal}
  {\bibinfo  {journal} {AIP Conf.Proc.}\ }\textbf {\bibinfo {volume} {1200}},\
  \bibinfo {pages} {131} (\bibinfo {year} {2010})},\ \Eprint
  {http://arxiv.org/abs/0912.5375} {arXiv:0912.5375 [hep-ph]} \BibitemShut
  {NoStop}%
\bibitem [{\citenamefont {Kolesova}\ and\ \citenamefont
  {Malinsky}(2014)}]{Kolesova:2014mfa}%
  \BibitemOpen
  \bibfield  {author} {\bibinfo {author} {\bibfnamefont {H.}~\bibnamefont
  {Kolesova}}\ and\ \bibinfo {author} {\bibfnamefont {M.}~\bibnamefont
  {Malinsky}},\ }\href@noop {} {\  (\bibinfo {year} {2014})},\ \Eprint
  {http://arxiv.org/abs/1409.4961} {arXiv:1409.4961 [hep-ph]} \BibitemShut
  {NoStop}%
\bibitem [{\citenamefont {Babu}\ and\ \citenamefont
  {Mohapatra}(2012)}]{Babu:2012vc}%
  \BibitemOpen
  \bibfield  {author} {\bibinfo {author} {\bibfnamefont {K.}~\bibnamefont
  {Babu}}\ and\ \bibinfo {author} {\bibfnamefont {R.}~\bibnamefont
  {Mohapatra}},\ }\href {\doibase 10.1016/j.physletb.2012.08.006} {\bibfield
  {journal} {\bibinfo  {journal} {Phys.Lett.}\ }\textbf {\bibinfo {volume}
  {B715}},\ \bibinfo {pages} {328} (\bibinfo {year} {2012})},\ \Eprint
  {http://arxiv.org/abs/1206.5701} {arXiv:1206.5701 [hep-ph]} \BibitemShut
  {NoStop}%
\bibitem [{\citenamefont {An}\ \emph {et~al.}(2010)\citenamefont {An},
  \citenamefont {Chen}, \citenamefont {Mohapatra},\ and\ \citenamefont
  {Zhang}}]{An:2009vq}%
  \BibitemOpen
  \bibfield  {author} {\bibinfo {author} {\bibfnamefont {H.}~\bibnamefont
  {An}}, \bibinfo {author} {\bibfnamefont {S.-L.}\ \bibnamefont {Chen}},
  \bibinfo {author} {\bibfnamefont {R.~N.}\ \bibnamefont {Mohapatra}}, \ and\
  \bibinfo {author} {\bibfnamefont {Y.}~\bibnamefont {Zhang}},\ }\href
  {\doibase 10.1007/JHEP03(2010)124} {\bibfield  {journal} {\bibinfo  {journal}
  {JHEP}\ }\textbf {\bibinfo {volume} {1003}},\ \bibinfo {pages} {124}
  (\bibinfo {year} {2010})},\ \Eprint {http://arxiv.org/abs/0911.4463}
  {arXiv:0911.4463 [hep-ph]} \BibitemShut {NoStop}%
\bibitem [{\citenamefont {Markevitch}\ \emph {et~al.}(2004)\citenamefont
  {Markevitch}, \citenamefont {Gonzalez}, \citenamefont {Clowe}, \citenamefont
  {Vikhlinin}, \citenamefont {David} \emph {et~al.}}]{Markevitch:2003at}%
  \BibitemOpen
  \bibfield  {author} {\bibinfo {author} {\bibfnamefont {M.}~\bibnamefont
  {Markevitch}}, \bibinfo {author} {\bibfnamefont {A.}~\bibnamefont
  {Gonzalez}}, \bibinfo {author} {\bibfnamefont {D.}~\bibnamefont {Clowe}},
  \bibinfo {author} {\bibfnamefont {A.}~\bibnamefont {Vikhlinin}}, \bibinfo
  {author} {\bibfnamefont {L.}~\bibnamefont {David}},  \emph {et~al.},\ }\href
  {\doibase 10.1086/383178} {\bibfield  {journal} {\bibinfo  {journal}
  {Astrophys.J.}\ }\textbf {\bibinfo {volume} {606}},\ \bibinfo {pages} {819}
  (\bibinfo {year} {2004})},\ \Eprint {http://arxiv.org/abs/astro-ph/0309303}
  {arXiv:astro-ph/0309303 [astro-ph]} \BibitemShut {NoStop}%
\bibitem [{\citenamefont {Spergel}\ and\ \citenamefont
  {Steinhardt}(2000)}]{Spergel:1999mh}%
  \BibitemOpen
  \bibfield  {author} {\bibinfo {author} {\bibfnamefont {D.~N.}\ \bibnamefont
  {Spergel}}\ and\ \bibinfo {author} {\bibfnamefont {P.~J.}\ \bibnamefont
  {Steinhardt}},\ }\href {\doibase 10.1103/PhysRevLett.84.3760} {\bibfield
  {journal} {\bibinfo  {journal} {Phys.Rev.Lett.}\ }\textbf {\bibinfo {volume}
  {84}},\ \bibinfo {pages} {3760} (\bibinfo {year} {2000})},\ \Eprint
  {http://arxiv.org/abs/astro-ph/9909386} {arXiv:astro-ph/9909386 [astro-ph]}
  \BibitemShut {NoStop}%
\end{thebibliography}%


\end{document}